\documentclass[%
 reprint,
 amsmath,amssymb,
 aps,
]{revtex4-2}

\usepackage{graphicx}% Include figure files
\usepackage{dcolumn}% Align table columns on decimal point
\usepackage{bm}% bold math
\usepackage[utf8]{inputenc}
\usepackage[T1]{fontenc}
\usepackage{hyperref}
\usepackage{amssymb}
\usepackage{amsmath}
\usepackage[scr=rsfso,cal=zapfc,frak=euler,bb=ams]{mathalfa}
\usepackage{caption}
\usepackage{subcaption}
\usepackage{tikz}
\usepackage{xcolor}

\DeclareMathAlphabet\mathbfcal{OMS}{cmsy}{b}{n}

\begin{document}

\preprint{APS/123-QED}

\title{Stability and decoherence analysis\\ of the silicon vacancy in 3C-SiC}% Force line breaks with \\

\author{Tommaso Fazio$^{1,2,5}$}
 \altaffiliation[]{tommaso.fazio@unipa.it}
\author{Ioannis Deretzis$^5$}
 \altaffiliation[]{ioannis.deretzis@imm.cnr.it}
\author{Giuseppe Fisicaro$^5$}
\author{Elisabetta Paladino$^{2,3,4}$}
\author{Antonino La Magna$^5$}

\affiliation{%
 $^1$Dipartimento di Fisica e Chimica "Emilio Segré", Università degli Studi di Palermo, Via Archirafi 36,\\ 90123 Palermo, Italy
}%

\affiliation{%
 $^2$Dipartimento di Fisica e Astronomia ”Ettore Majorana”, Università di Catania, Via S. Sofia 64, 95123 Catania, Italy
}%
\affiliation{
 $^3$CNR-IMM, Catania (University unit), Consiglio Nazionale delle Ricerche,\\ Via S. Sofia 64, 95123 Catania, Italy
}%
\affiliation{
 $^4$Istituto Nazionale di Fisica Nucleare, Sezione di Catania,\\ Via S. Sofia 64, 95123 Catania, Italy
}%

\affiliation{
 $^5$Consiglio Nazionale delle Ricerche, Istituto per la Microelettronica e Microsistemi (CNR-IMM), Z.I. VIII Strada 5, 95121 Catania, Italy
}%


\date{\today}% It is always \today, today,

\begin{abstract}
\noindent The silicon vacancy ($\mathrm{V}_{\mathrm{Si}}$) in 3C-SiC is studied as a center of interest in the field of Quantum Technologies, modeled as an electron spin (behaving as a two state qubit in appropriate conditions) interacting through hyperfine coupling with the SiC nuclear spin bath containing $^{29}\mathrm{Si}$ and $^{13}\mathrm{C}$ nuclei in their natural isotopic concentration. We calculate the formation energies of the neutral and charged $\mathrm{V}_{\mathrm{Si}}$ with \textit{ab initio} methods based on the Density Functional Theory, identifying the stability of the neutral charge state for energies close to the valence band of 3C-SiC. In addition, magnetic properties are calculated for the $\mathrm{V}_{\mathrm{Si}}^{-1}$ in 3C-SiC and for $\mathrm{V}_{\mathrm{Si}}^0$ in both cubic and hexagonal SiC polytypes. We thereon evaluate, for the defect in the cubic polytype, the Free Induction Decay and the Hahn-echo sequence on the electron spin interacting with the nuclear spin bath, shedding light on the Electron Spin Echo Envelope Modulation phenomenon and the decoherence effect by means of the Cluster Correlation Expansion theory. We find a non-exponential coherence decay, which is a typical feature of solid-state qubits subjected to low frequency 1/f-type noise from the environment. 
%\begin{description}
%\end{description}
\end{abstract}

\maketitle

%\tableofcontents

\section{\label{sec:intro}Introduction}%\protect\\

Silicon carbide (SiC) is widely recognized as an interesting material for technological applications. Its capacity to work in harsh environments under high temperatures, yielding faster switching speeds, lower power losses and higher blocking voltages with respect to silicon has boosted its industrial exploitation in solid-state devices~\cite{kimoto2014fundamentals}. Within this context, SiC-based architectures have been already utilized, for instance, in accelerator physics~\cite{okamura2017high} and in microelectronics as power devices~\cite{alves2017sic}. Recently, it has gained popularity in the field of Quantum Technologies (QT), e.g. used as a sensor of magnetic fields and temperature gradients~\cite{castelletto2020silicon}. Here, point-like defective configurations like divacancies or $\mathrm{Si}/\mathrm{C}$ vacancies can provide active states on which quantum information is encoded, processed and stored. Out of the many SiC polytypes, where the difference lies in the structure of the stacking layers, the most studied are the hexagonal 4H- and 6H-SiC~\cite{ivady2015theoretical} ones, due to the significant progress made in their epitaxial growth and the easy access to samples having low concentrations of defects. On the other hand, the cubic 3C-SiC polytype could potentially be an interesting and convenient alternative due to the possibility to be heteroepitaxially grown on silicon substrates, along with a series of physical characteristics which are appealing for electronic devices (lower band gap, absence of deep level stacking-fault defect states, higher electron and hole mobilities, etc.)~\cite{senesky2009harsh,fisicaro2020genesis}. However, the low quality of 3C-SiC crystals during the past has largely hindered its technological use and slowed down its further theoretical study. Recently, new and improved fabrication techniques have been introduced that lower the defect concentration in cubic 3C-SiC during growth~\cite{fisicaro2020genesis,la2021new}. Therefore, the cubic SiC polytype could gather the interest of the QT community, as previously happened for the 6H- and 4H-SiC polytypes~\cite{seo2016quantum}. \\ 
\hspace*{0.5cm}The availability of high-quality and precisely-doped 3C-SiC samples could aid the definitive assessment of single vacancy-related centers for possible QT applications in this material. Indeed, contrarily to other vacancy-related centers as the di-vacancy or the nitrogen-vacancy complex~\cite{gordon2015defects}, the single vacancy in the 3C-SiC material has not been unambiguously characterized in experiments so far in terms of optical and magnetic signatures. We notice that the presence of a high density of stacking fault defects in conventional hetero-epitaxial 3C-SiC layers makes it difficult to identify unknown infrared emitters which show the characteristics of intrinsic defect-related color centers in 3C-SiC, but with different optical features with respect to the di-vacancy one (see refs.~\cite{wang2018bright,castelletto2020silicon}. Moreover, this difficult assignment makes also a topic of debate the conclusion of ref.~\cite{lefevre2011characterization}, where the detected $T_X$ electron paramagnetic resonance (EPR) signal in 3C-SiC has been tentatively associated to the silicon vacancy in its neutral state, but in an n-type specimen where the charged configurations should be the most abundant vacancy states~\cite{torpo1999silicon,kawasuso1998silicon}. \\
\hspace*{0.5cm}In view of the emerging new paradigms for the 3C-SiC material application, we theoretically address the polarity dependent states of the silicon vacancy in 3C-SiC which could have a strong potentiality for QTs due to the low rate of quantum information loss that vacancy related states exhibit in SiC ~\cite{seo2016quantum,fazio2022computational}, even at room temperature~\cite{falk2013polytype}. We mainly focus on the neutral S=1 ($\mathrm{V}_{\mathrm{Si}}^0$) state, but the extension of the full approach to the S=3/2 ($\mathrm{V}_{\mathrm{Si}}^{-1}$) is also considered. To reproduce the local environment of a non-defective 3C-SiC bulk region we assume that the defect is coupled through hyperfine interactions with the SiC nuclear spin bath, constituted by naturally occurring $^{29}\mathrm{Si}$ and $^{13}\mathrm{C}$ paramagnetic nuclei. We, indeed, notice that the time evolution of the spin center correlated to the defect is affected by the magnetic environment of the $^{29}\mathrm{Si}$ and $^{13}\mathrm{C}$ nuclear spins-1/2. Moreover, microwave pulses~\cite{ivady2015theoretical} tuned to a particular transition frequency can be used to control the defect and reduce its relevant eigenstates entering the dynamics to two, thereby effectively dealing with a qubit. This study is complemented by the identification of the stability window in terms of electrochemical potential comparing the computed formation energy with those of charged silicon vacancies. The energetics~\cite{giannozzi2009quantum} and hyperfine interaction~\cite{pickard2001all} properties of these defects can be calculated via \textit{ab initio} methods based on the Density Functional Theory (DFT). We note that this vacancy state has received less attention in the theoretical literature with respect to other defect-impurity states in SiC. \\ 
\hspace*{0.5cm}The nuclear spin bath induces noise at low frequencies for the point defect. This is typically the case in Nuclear Magnetic Resonance (NMR), where the interesting experimental signal is generated by nonequilibrium electron spin magnetization (equivalent to its coherence) precessing about an external magnetic field~\cite{slichter2013principles}. Due to the spatial field inhomogeneity, the measured signal in a Free-Induction Decay (FID) process is defocused and displays a characteristic non-exponential decay resulting in inhomogeneous broadening of the spectral lines. Analogous effects occur in solid-state implementations of qubits and originate from time-inhomogeneities due to repetitions of measurement protocols~\cite{paladino20141,falci2005initial,bergli2009decoherence}. The same happens when the degree of freedom of the examined system is electronic in nature, as opposed to nuclear. This is the case in Electron Paramagnetic Resonance (EPR), in which one applies NMR techniques to an electron spin~\cite{al2013electron}. One way of refocusing can be achieved by the Hahn-echo sequence, an established technique applied recently to investigate the residual decoherence of divacancy defects in 4H-SiC in ref.~\cite{seo2016quantum}. Here we consider both the FID and Hahn-echo sequences applied to the considered defect utilizing the Cluster Correlation Expansion (CCE) theory~\cite{yang2008quantum}. CCE theory allows to split the bath in clusters with a given number of interacting nuclear spins. Since the clusters by hypothesis are \textit{uncorrelated} with each other, the qubit coherence is obtained as a product of the contributions of each cluster. \\
\hspace*{0.5cm}The rest of the paper is organized as follows: in section 2 we report our results by describing the model used and the \textit{ab initio} calculations for the $\mathrm{V}_{\mathrm{Si}}^0$, $\mathrm{V}_{\mathrm{Si}}^{-1}$ and $\mathrm{V}_{\mathrm{Si}}^{-2}$ in 3C-SiC, their formation energy as a function of the Fermi level and the calculation of the magnetic parameters of the system plus environment Hamiltonian, i.e. the hyperfine tensor and the Zero-Field Splitting (ZFS) tensor. At the end of the section we utilize the magnetic parameters derived from first principles to calculate both analytically and numerically the spin coherence (or its decay) after free evolution (FID) and under the Hahn-echo sequence, at a CCE1 and CCE2 level. In section 3 a discussion on our results is presented, while finally in section 4 our methods of analysis behind the \textit{ab initio} calculations and the evaluation of the qubit decoherence are described in detail.

\section{\label{sec:results}Results}

\subsection{\label{subsec:model}Model}

\begin{figure}
\centering
\begin{subfigure}{0.4944\linewidth}
\caption{}
\label{fig:spin density}
    \includegraphics[width=1.0\linewidth]{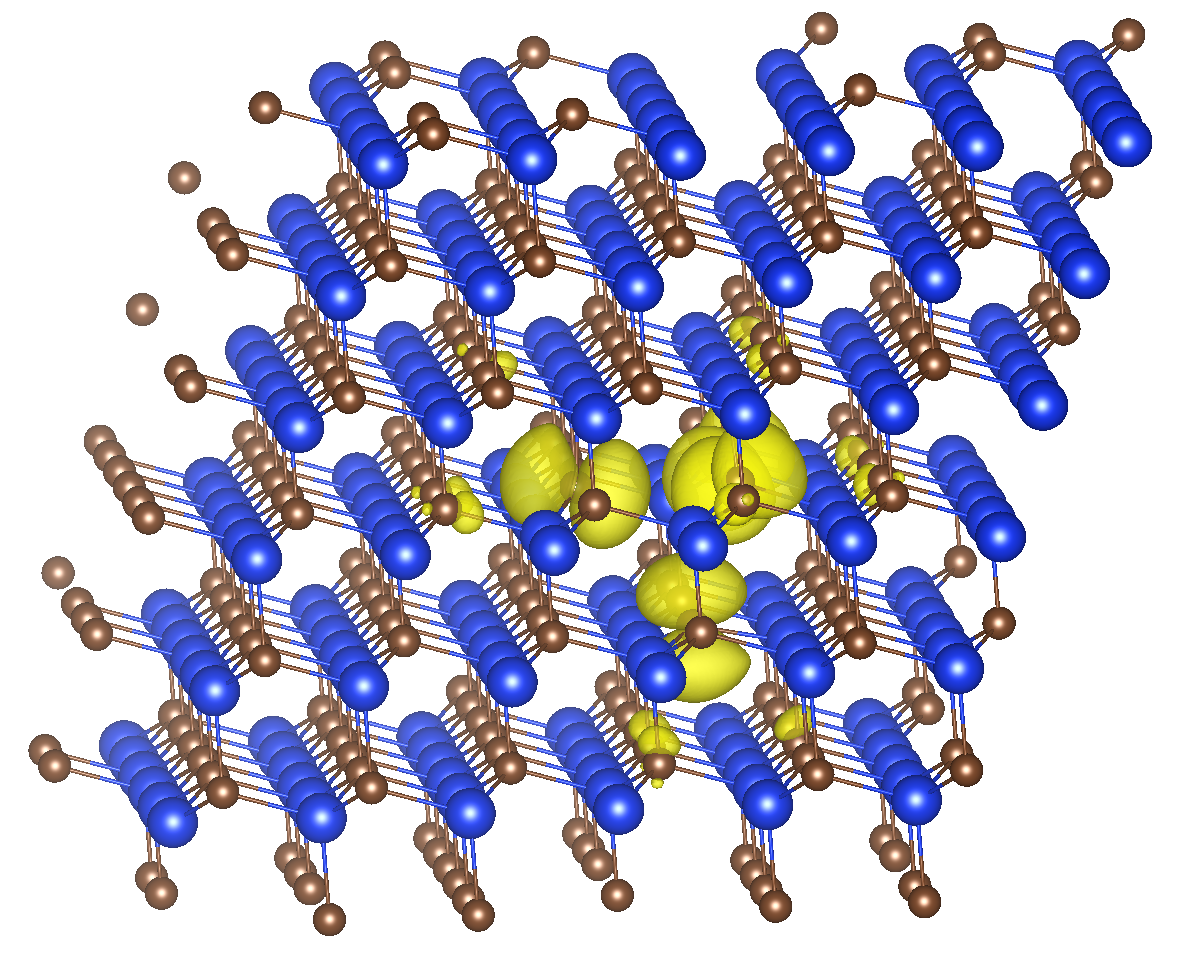}
\end{subfigure}
\hfill
\begin{subfigure}{0.4944\linewidth}
\caption{}
\label{fig:formation energies graph}
    \includegraphics[width=1.0\linewidth]{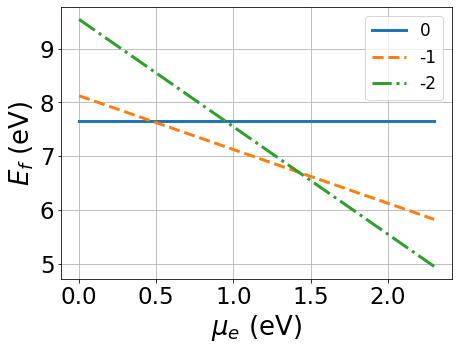}
\end{subfigure}
\caption{(a) 3D spin density around a neutral $\mathrm{V}_{\mathrm{Si}}$ in 3C-SiC. The spin density differences are mainly extended until the third neighbor shell. The wave functions are calculated for a $6\times6\times6$ 3C-SiC supercell. (b) Formation energies of a neutral, -1 and -2 charged $\mathrm{V}_{\mathrm{Si}}$ as a function of the Fermi level with respect to the valence band maximum, for non-collinear calculations in a 686-atom 3C-SiC supercell. The monopole term is included in the analysis.}
\label{fig:formation energies graph and spin density}
\end{figure}

\textit{Ab initio} calculations are useful for assessing structural, electronic, optical and magnetic properties of solids~\cite{chalasinski1994origins}. Here we use the density functional theory to evaluate the magnetic parameters of a Hamiltonian describing the interaction between the electron and nuclear spins through the calculation of the hyperfine ~\cite{szasz2013hyperfine} and the ZFS tensors~\cite{ivady2014pressure}. The following is our working Hamiltonian (the choice of putting $E=0$ is justified in Appendix \ref{app:A})~\cite{sakuldee2019characterization}
\begin{equation}
\begin{split}
\mathcal{H}&=D S_{z}^2+\gamma_{e}B_{z}S_{z}+\sum_{i=1}^N\gamma_{i}B_{z}I_{iz} \\
&+S_{z}\otimes\sum_{i=1}^N\left(A_{i}I_{iz}+B_{i}I_{ix}\right)+\mathcal{H}_{n-n}, \label{eq:hamtotpuredeph}
\end{split}
\end{equation}
where $\mathcal{H}_{n-n}$ is the dipolar interaction between nuclear spins, $A_{i}\equiv A_{zz}^i$ and $B_{i}\equiv(A_{zx}^{i^2}+A_{zy}^{i^2})^{1/2}$, whereas $A_{zx}$, $A_{zy}$ and $A_{zz}$ are the elements of the third row of the hyperfine tensor. Eq. \ref{eq:hamtotpuredeph} is already written in the pure-dephasing approximation~\cite{krummheuer2002theory,sakuldee2019characterization}, so that no transition of the electron spin takes place by exchanging energy with the environment. Note that the Hamiltonian \ref{eq:hamtotpuredeph} commutes with the electron spin $S_{z}$ operator and, if we consider with no lack of generality a $S=1$ system, it can be expressed in the spin operator eigenbasis $\lbrace\vert1\rangle,\vert0\rangle,\vert-1\rangle\rbrace$, giving rise to~\cite{seo2016quantum}
\begin{equation}
\mathcal{H}=\sum_{m_{S}=1,0,-1}\vert m_{S}\rangle\langle m_{S}\vert\otimes\mathcal{H}_{m_{S}}, \label{eq:hamtotpuredephcompact}
\end{equation}
where
\begin{equation}
\mathcal{H}_{m_{S}}=\omega_{m_{S}}+\mathcal{H}_{\mathrm{B}}+m_{S}\sum_{i=1}^N(A_{i}I_{iz}+B_{i}I_{ix}). \label{eq:hamtotpuredephcompactcond}
\end{equation}
Furthermore, $\mathcal{H}_{\mathrm{B}}=\sum_{i=1}^N\gamma_{i}B_{z}I_{iz}+\mathcal{H}_{n-n}$ is the bath Hamiltonian. Finally,
\begin{align}
\omega_{1}&=D+\omega_{e}, \label{eq:omega1} \\
\omega_{0}&=0, \label{eq:omega0} \\
\omega_{-1}&=D-\omega_{e}, \label{eq:omega-1}
\end{align}
where $\omega_{e}=\gamma_{e}B$ is the Larmor frequency of the electron spin, are the eigenvalues of the electron spin Hamiltonian (first and second term of Eq. \ref{eq:hamtotpuredeph}). A direct consequence of the form of Hamiltonian \ref{eq:hamtotpuredephcompact} is that, by opportunely initializing the electron spin (more on that in section 4) and appropriately choosing the control pulses as having precisely the right frequency $\omega_{1}$, the $\vert-1\rangle$ state can be frozen out of the dynamics since no transitions are allowed towards it. Therefore, the electron spin effectively behaves as a qubit~\cite{seo2016quantum}. \\
\hspace*{0.5cm}As a first-order approximation, in calculating the hyperfine tensor the electron and nuclear spins can be considered as point-dipoles, which is known in the literature as the semiclassical approximation~\cite{schweiger2001principles}. In the semiclassical approximation, the hyperfine tensor is given by
\begin{equation}
\mathbfcal{A}_{i}=\frac{\mu_{0}\gamma_{i}\gamma_{e}}{4\pi r_{i}^3}\left(1-\frac{3\textbf{r}_{i}\textbf{r}_{i}}{r_{i}^2}\right), \label{eq:smhyptens}
\end{equation}
where $\mu_{0}$ is the magnetic permeability of the vacuum, $\gamma_{i}$ and $\gamma_{e}$ are the $i$-th nuclear spin and electron spin gyromagnetic ratios, respectively, whereas $\textbf{r}_{i}$ is the position vector of the $i$-th nuclear spin with respect to the qubit, with its modulus $r_{i}$ being the distance between the two. Of course Eq. \ref{eq:smhyptens} is no longer applicable in the immediate vicinity of the qubit. \textit{Ab initio} methods allow us to go beyond the semiclassical approximation and model physical effects generated by the three-dimensional defect's spin density (see Fig. \ref{fig:spin density}). \\

\subsection{\label{subsec:form_ener}Formation energy}

An important issue for the determination of the stability of a particular defect under given thermodynamic conditions regards the energetic competition between its various charged states. Here we have used the DFT to calculate the formation energy of the neutral and charged $\mathrm{V}_{\mathrm{Si}}$ within a 3C-SiC $7\times7\times7$ supercell. The formation energy $E_{f}$ of a defect $X$ can be defined as the energy difference between the investigated system and the components in their reference states~\cite{freysoldt2014first}, i.e.
\begin{equation}
\begin{split}
E_{f}\left[X^q\right]&=E_{\mathrm{tot}}\left[X^q\right]-E_{\mathrm{tot}}\left[\mathrm{bulk}\right]-\sum_{i}n_{i}\mu_{i} \\
&+\frac{q}{e}\left(E_{\mathrm{VBM}}+\mu_{e}\right)+E_{\mathrm{corr}}. \label{eq:form_en}
\end{split}
\end{equation}
$E_{\mathrm{tot}}\left[X^q\right]$ is the total energy of the host crystal with the defect with charge $q$, where $e>0$ is the elementary charge of the electron, $E_{\mathrm{tot}}\left[\mathrm{bulk}\right]$ is the total energy of the same cell of crystal without the defect, and $n_{i}\mu_{i}$ is the reference energy of added (or subtracted with a change of sign) atoms of element $i$ at chemical potential $\mu_{i}$. The term in parenthesis accounts for the chemical potential of the electron(s) involved in charging the defect. $E_{\mathrm{VBM}}$ is the valence band maximum as given by the calculated band structure for the bulk material, and $\mu_{e}$ is the electron chemical potential defined here with respect to the top of the corresponding valence band. The $\mu_{e}$ parameter can then be treated as a free parameter, allowing to account for a shift of the Fermi level, e.g., due to doping. Note that $\mu_{e}=E_{\mathrm{gap}}/2$ corresponds to the undoped semiconductor case, where $E_{\mathrm{gap}}$ is the intrinsic semiconductor band gap. Finally, $E_{\mathrm{corr}}$ is a sum of relevant correction terms, the most important of which is the monopole correction term, taking into account the electrostatic interaction between the charged defect and its periodic replicas within the \textit{ab initio} simulations. The monopole correction term can be written as~\cite{bockstedte2003ab}
\begin{equation}
E_{\mathrm{corr}}=\frac{q^2\alpha}{2\epsilon L}, \label{eq:form_en_corr}
\end{equation}
where $q$ is the charge of the defect, $\alpha$ is the Madelung constant associated to our crystal structure, $\epsilon$ is the SiC experimental dielectric constant and $L$ is the distance between the defect and its periodic replicas. For the neutral $\mathrm{V}_{\mathrm{Si}}$ we have calculated $E_{\mathrm{tot}}\left[\mathrm{V}_{\mathrm{Si}}^0\right]-E_{\mathrm{tot}}\left[\mathrm{bulk}\right]$ and $\mu_{\mathrm{Si}}$, whereas for the charged defects, for which $q$ is different from zero, we have also calculated the valence band maximum (see Eq. \ref{eq:form_en}). All \textit{ab initio} calculations are based on the DFT by using the Perdew-Burke-Ezernhof implementation~\cite{perdew1996generalized} of the generalized gradient approximation for the description of the exchange-correlation functional, as implemented in the Quantum Espresso (QE) software suite~\cite{giannozzi2009quantum}. (see the Methods section for more details). We note here that the focus of our analysis lies on the magnetic properties of the neutral and charged silicon vacancy, whereas, for a more accurate interpretation of the electronic structure, approaches like the many-body perturbation theory or well-selected hybrid density functional theory could be useful. \\
\hspace*{0.5cm}Upon structural relaxation which induces a local reconstruction around the defected site~\cite{oda2013study}, the calculated magnetization for the $\mathrm{V}_{\mathrm{Si}}^0$, $\mathrm{V}_{\mathrm{Si}}^{-1}$ and $\mathrm{V}_{\mathrm{Si}}^{-2}$ defects was the one expected for a defect with electron spin-1, 3/2 and 1, respectively~\cite{torpo1999silicon}. Fig. \ref{fig:spin density} shows the spin density around the central $\mathrm{V}_{\mathrm{Si}}$, which extends until the third neighbor shell. This nonzero spin density is modeled and implemented in the QE-GIPAW code~\cite{varini2013enhancement} and allows us to go beyond the semiclassical magnetic point-dipole approximation of Eq. \ref{eq:smhyptens} (see the next section). \\
\hspace*{0.5cm}As we can see in Fig. \ref{fig:formation energies graph}, in which the formation energy of a $\mathrm{V}_{\mathrm{Si}}$ with different charge states is shown as a function of the Fermi level with respect to the valence band maximum (corrected with the monopole correction term given in Eq. \ref{eq:form_en_corr}), our \textit{ab initio} results demonstrate the stability of the neutral state near the valence band of 3C-SiC. Moreover, the presence of the monopole term widens the electrochemical potential range over which the neutral state is stable. These results are in good agreement with previous studies~\cite{xi2016ab,oda2013study,bockstedte2003ab}. Hence, the neutral silicon vacancy in 3C-SiC could potentially be stable in $p$-doped 3C-SiC systems, as for example in Al-doped 3C-SiC three gate MOSFET devices~\cite{schoner2006fabrication}. Within this context, in the next section the hyperfine interactions for a $\mathrm{V}_{\mathrm{Si}}^0$ and $\mathrm{V}_{\mathrm{Si}}^{-1}$ in 3C-SiC as well as for the neutral defect in 4H-SiC are calculated from first principles, noting that the $\mathrm{V}_{\mathrm{Si}}^{-1}$ state in 4H-SiC has been thoroughly studied in the literature~\cite{ivady2017identification,bockstedte2003signature,isoya2008epr,lefevre2011characterization}. Moreover, the ZFS tensor components of  $\mathrm{V}_{\mathrm{Si}}^0$ and $\mathrm{V}_{\mathrm{Si}}^{-1}$ in 3C-SiC are also computed. \\

\begin{table}
    \centering
    \begin{tabular}{|p{2cm}|p{1cm}|p{1cm}|p{1cm}|}
      \hline
      atom & $A_{xx}$ & $A_{yy}$ & $A_{zz}$ \\
      \hline
      $\hspace{0.3cm}\mathrm{C}_{1}$ & 27.4 & 27.4 & 85.6 \\
      \hline
      $\hspace{0.3cm}\mathrm{C}_{2}$ & 27.4 & 27.4 & 85.4 \\
      \hline
      $\hspace{0.3cm}\mathrm{C}_{3}$ & 27.6 & 27.6 & 85.8 \\
      \hline
      $\hspace{0.3cm}\mathrm{C}_{4}$ & 27.4 & 27.4 & 85.6 \\
      \hline
      $\hspace{0.3cm}\mathrm{Si}_{1}-\mathrm{Si}_{12}$ & 7.4 & 7.6 & 6.9 \\
      \hline
    \end{tabular}
    \caption{\textit{Ab initio} calculated values (in $\mathrm{MHz}$) for the hyperfine tensor components describing the interaction between a neutral $\mathrm{V}_{\mathrm{Si}}$ in 3C-SiC and the nuclear spins in the first and second neighbor shells. %The hyperfine matrices are digonalized and the angles between the crystallographic axis and the principal axis of diagonalization for the C$_1$ and Si atoms are 54.7$^\circ$ and 111.7$^\circ$, respectively. 
    The values are obtained by using the QE-GIPAW code~\cite{varini2013enhancement}.}
    \label{tab:my_label}
\end{table}

\begin{table}
    \centering
    \begin{tabular}{|p{2cm}|p{1cm}|p{1cm}|p{1cm}|}
      \hline
      atom & $A_{xx}$ & $A_{yy}$ & $A_{zz}$ \\
      \hline
      $\hspace{0.3cm}\mathrm{C}_{1}$ & 27.2 & 27.2 & 79.7 \\
      \hline
      $\hspace{0.3cm}\mathrm{C}_{2}$ & 27.2 & 27.2 & 79.7 \\
      \hline
      $\hspace{0.3cm}\mathrm{C}_{3}$ & 27.2 & 27.2 & 79.7 \\
      \hline
      $\hspace{0.3cm}\mathrm{C}_{4}$ & 27.2 & 27.2 & 79.7 \\
      \hline
      $\hspace{0.3cm}\mathrm{Si}_{1}-\mathrm{Si}_{12}$ & 7.4 & 7.5 & 6.8 \\
      \hline
    \end{tabular}
    \caption{\textit{Ab initio} calculated values (in $\mathrm{MHz}$) for the hyperfine tensor components describing the interaction between a negatively charged $\mathrm{V}_{\mathrm{Si}}$ in 3C-SiC and the nuclear spins in the first and second neighbor shells. %The hyperfine matrices are digonalized and the angles between the crystallographic axis and the principal axis of diagonalization for the C$_1$ and Si atoms are 54.7$^\circ$ and 111.7$^\circ$, respectively. 
    The values are obtained by using the QE-GIPAW code~\cite{varini2013enhancement}.}
    \label{tab:my_label_1}
\end{table}

\subsection{\label{subsec:hyp_int_zfs}Hyperfine interactions and Zero Field splitting}

The results of our \textit{ab initio} calculations can be used to define the hyperfine and ZFS tensors from first principles with the aid of the QE-GIPAW~\cite{varini2013enhancement} and PyZFS~\cite{ma2020pyzfs} codes, respectively. Table \ref{tab:my_label} and Table \ref{tab:my_label_1} show the hyperfine tensor components describing the interaction between a neutral or a negatively charged $\mathrm{V}_{\mathrm{Si}}$ in 3C-SiC and the nuclear spins in the first and second neighbor shells. In the case of the $\mathrm{V}_{\mathrm{Si}}^{-1}$, results are in good agreement with the experimental measurements of Ref.~\cite{itoh1997intrinsic}. Tables \ref{tab:my_label1} and \ref{tab:my_label2} show the same components for a neutral $\mathrm{V}_{\mathrm{Si}}$ in 4H-SiC, located in the two nonequivalent 4H sites (i.e., $k$ and $h$), respectively. In this case, a comparison with the experimental results of Ref.~\cite{wagner2002ligand} seems also reasonable. 

Concerning the ZFS tensor components of the electron spin associated to a $\mathrm{V}_{\mathrm{Si}}^0$ in 3C-SiC, the values obtained for the axial and transversal components are $D=0.45\hspace{0.1cm}\mathrm{MHz}$ and $E=-0.09\hspace{0.1cm}\mathrm{MHz}$, respectively. The tensor components values for the charged defect $\mathrm{V}_{\mathrm{Si}}^{-1}$ instead are $D=0.1\hspace{0.1cm}\mathrm{MHz}$ and $E=-0.03\hspace{0.1cm}\mathrm{MHz}$. As we can see, the absolute values of the components are relatively low and they seem to indicate the presence of axial symmetry for the ZFS tensor. This allows us to predict an apparent $T_d$ symmetry of the $\mathrm{V}_{\mathrm{Si}}^0$ $a_1$ and $e$ states, which is in agreement with the finding in Lefévre et al.~\cite{lefevre2011characterization} of an axial-symmetric ZFS tensor for the same defect in 3C-SiC. Anyhow, we notice that the attribution in ref.~\cite{lefevre2011characterization} appears in an n-doped material where our calculations show the stability of charged defects. Concerning the symmetry of the $\mathrm{V}_{\mathrm{Si}}^0$ ground state in 3C-SiC, there is no unanimous consensus in the literature, whereas the spin state is almost certain. Early works utilizing small supercells assigned to the $\mathrm{V}_{\mathrm{Si}}^0$ a $^1E$ spin singlet configuration~\cite{deak1999spin,oda2013study} or a $^3T_1$ spin triplet configuration with $T_d$ symmetry~\cite{zywietz2000spin}. A recent work with larger supercells identifies, as in our case, a spin triplet configuration, but with a Jahn-Teller-distorted $C_{3v}$ symmetry as the ground state~\cite{schultz2021theoretical}. A further breakdown of the triplet state into a finer structure of levels is also probable, especially in the presence of dynamic polaronic distortion. Within this framework, we cannot exclude a multiplet nature of all the electronic configurations. However, as we thoroughly demonstrate in Appendix \ref{app:A}, a small variation of the ZFS tensor components has a minor influence on the electron spin dynamics. \\

\begin{table}
    \centering
    \begin{tabular}{|p{2cm}|p{1cm}|p{1cm}|p{1cm}|}
      \hline
      atom & $A_{xx}$ & $A_{yy}$ & $A_{zz}$ \\
      \hline
      $\hspace{0.3cm}\mathrm{C}_{1}$ & 24.5 & 24.4 & 76.3 \\
      \hline
      $\hspace{0.3cm}\mathrm{C}_{2}$ & 24.5 & 24.4 & 76.2 \\
      \hline
      $\hspace{0.3cm}\mathrm{C}_{3}$ & 34.4 & 34.4 & 110.4 \\
      \hline
      $\hspace{0.3cm}\mathrm{C}_{4}$ & 24.5 & 24.4 & 76.2 \\
      \hline
      $\hspace{0.3cm}\mathrm{Si}_{1}-\mathrm{Si}_{12}$ & 7.5 & 7.8 & 6.9 \\
      \hline
    \end{tabular}
    \caption{\textit{Ab initio} calculated values (in $\mathrm{MHz}$) for the hyperfine tensor components describing the interaction between a neutral $\mathrm{V}_{\mathrm{Si}}$ (k site) in 4H-SiC and the nuclear spins in the first and second neighbor shells. %The hyperfine matrices are digonalized and the angles between the crystallographic axis and the principal axis of diagonalization for the C$_1$ and Si atoms are 69.2$^\circ$ and 98.5$^\circ$, respectively. 
    The values are obtained by using the QE-GIPAW code~\cite{varini2013enhancement}.}
    \label{tab:my_label1}
\end{table}
\begin{table}
    \centering
    \begin{tabular}{|p{2cm}|p{1cm}|p{1cm}|p{1cm}|}
      \hline
      atom & $A_{xx}$ & $A_{yy}$ & $A_{zz}$ \\
      \hline
      $\hspace{0.3cm}\mathrm{C}_{1}$ & 24.3 & 24.2 & 75.1 \\
      \hline
      $\hspace{0.3cm}\mathrm{C}_{2}$ & 34.2 & 34.2 & 112.3 \\
      \hline
      $\hspace{0.3cm}\mathrm{C}_{3}$ & 24.4 & 24.3 & 76.0 \\
      \hline
      $\hspace{0.3cm}\mathrm{C}_{4}$ & 24.7 & 24.6 & 77.1 \\
      \hline
      $\hspace{0.3cm}\mathrm{Si}_{1}-\mathrm{Si}_{12}$ & 7.5 & 7.8 & 6.9 \\
      \hline
    \end{tabular}
    \caption{\textit{Ab initio} calculated values (in $\mathrm{MHz}$) for the hyperfine tensor components describing the interaction between a neutral $\mathrm{V}_{\mathrm{Si}}$ (h site) in 4H-SiC and the nuclear spins in the first and second neighbor shells. %The hyperfine matrices are digonalized and the angles between the crystallographic axis and the principal axis of diagonalization for the C$_1$ and Si atoms are 109.6$^\circ$ and 118.2$^\circ$, respectively. 
    The values are obtained by using the QE-GIPAW code~\cite{varini2013enhancement}.}
    \label{tab:my_label2}
\end{table}

\subsection{\label{subsec:fid}Free Induction Decay}

The research on FID is interesting for many reasons. In the literature, experiments are described that elucidate the quantum mechanical origins of the FID signal and spin noise~\cite{hoult2001quantum}. FID has also been used as a means of controlling the phase and amplitude of extreme ultraviolet photons~\cite{bengtsson2017space}. Our main objective in studying the FID, and the goal of this subsection, is to evaluate the qubit's decoherence time after free evolution and to compare it with the one obtained after a given control procedure is applied (cfr. Hahn-echo below). \\
\hspace*{0.5cm}Therefore, in this subsection we focus on the FID process~\cite{schweiger2001principles}, i.e. we let the system freely evolve after the preparation of the qubit. In general, the qubit coherence is defined as the off-diagonal component of the density matrix, or~\cite{seo2016quantum}
\begin{equation}
\mathcal{L}(t)\equiv\frac{\mathrm{tr}\left\lbrace\rho_{\mathrm{tot}}(t)S_{+}\right\rbrace}{\mathrm{tr}\left\lbrace\rho_{\mathrm{tot}}(0)S_{+}\right\rbrace}, \label{eq:totcoh}    
\end{equation}
where $\rho_{\mathrm{tot}}(t)$ is the total qubit plus bath density operator at time $t$, $S_{+}=S_{x}+iS_{y}$ is the qubit raising operator and $\rho_{\mathrm{tot}}(0)=\rho_{\mathrm{S}}(0)\otimes\rho_{\mathrm{B}}(0)$ is the initial state of the overall system. The spin bath is assumed in a totally mixed state. In the dipolar approximation of Eq. \ref{eq:hamtotpuredeph} the qubit eigenbasis coincides with a subset of the $S_{z}$ spin operator eigenbasis, i.e. $\lbrace\vert1\rangle,\vert0\rangle\rbrace$. The qubit is prepared in the pure state $\rho_{\mathrm{S}}(0)=\vert\Psi\rangle\langle\Psi\vert$, where
\begin{equation}
\vert\Psi\rangle=\frac{1}{\sqrt{2}}\big(\vert1\rangle+i\vert0\rangle\big), \label{eq:initstate}
\end{equation}
so that $\langle S_{y}\rangle(0)\neq 0$ and $\langle S_{x}\rangle(0)=0$. The preparation in state \ref{eq:initstate}, obtained via the application of a $\pi/2$ pulse to the qubit in the $\vert0\rangle$ state, together with the chosen form for the control pulses and the pure-dephasing approximation in Hamiltonian \ref{eq:hamtotpuredeph}, ensures that the $\vert-1\rangle$ state stays out of the dynamics. If suitable energy distribution of excited spin states occurs, in an eventual experimental realization of this protocol on a $\mathrm{V}_{\mathrm{Si}}^0$, the preparation in the $\vert0\rangle$ state could be achieved as in ref.~\cite{nagy2019high}, where a high fidelity in the initialization of the ground state for a $\mathrm{V}_{\mathrm{Si}}^{-1}$ in 4H-SiC has been demonstrated. It is known in the literature that this defect has a relatively low ZFS of $4$ MHz~\cite{biktagirov2018polytypism}. Then, different resonant energies are required for $\vert0\rangle$ and $\vert \pm 1 \rangle$ for the optical transitions. This is true for a $\mathrm{V}_{\mathrm{Si}}^{-1}$ in 4H-SiC~\cite{nagy2019high} and we cautiously suggest that differences with respect to a $\mathrm{V}_{\mathrm{Si}}^0$ in 3C-SiC should not be stark. This stage, or other methods for unbalanced selective transitions for $\vert0\rangle$ and $\vert \pm 1 \rangle$ states, will be reached whenever the defect's signatures in 3C-SiC will be detected. Here we provide a practical explanation, based on analytical calculations, of why it is useful to consider the $\mathrm{V}_{\mathrm{Si}}^0$ as a qubit. The coherence $\mathcal{L}(t)$ is a complex function having the expectation values of the qubit $S_{x}$ and $S_{y}$ operators as real and imaginary parts, respectively. Furthermore, Eq. \ref{eq:totcoh} becomes intractable rather quickly as the number of nuclear spins in the bath increases. The objective of CCE theory is then to provide a reasonable and computationally achievable approximated version of the whole coherence given in Eq. \ref{eq:totcoh}. In order to do so, the first step is the implementation of a numerical procedure generating a random bath of nuclear spins. The $^{29}\mathrm{Si}$ and $^{13}\mathrm{C}$ nuclear spins are thereby randomly put in our simulated 3C-SiC lattice, by using a random number generator, according to their natural abundance of $4.7$ and $1.1\%$, respectively. \\
\hspace*{0.5cm}The entire FID process can be described as
\begin{equation}
\rho_{\mathrm{FID}}(\tau)=U_{\mathrm{FID}}(\tau)\rho_{\mathrm{S}}(0)U_{\mathrm{FID}}^\dagger(\tau), \label{eq:fidprocess}    
\end{equation}
\begin{figure}
\centering
\begin{subfigure}{0.4944\linewidth}
\caption{}
\label{fig:FID CCE1-2 200G}
    \includegraphics[width=1.0\linewidth]{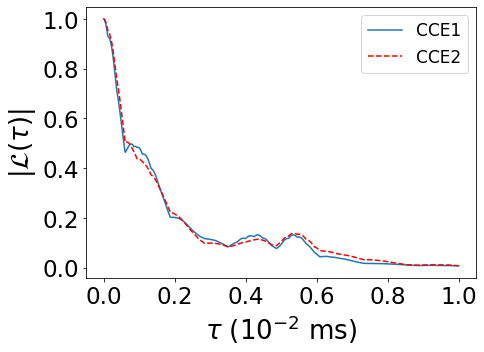}
\end{subfigure}
\hfill
\begin{subfigure}{0.4944\linewidth}
\caption{}
\label{fig:FID CCE1-2 500G}
    \includegraphics[width=1.0\linewidth]{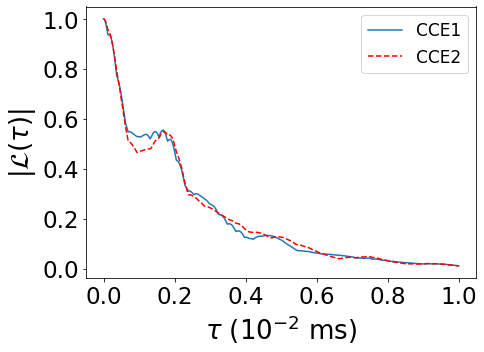}
\end{subfigure}
\caption{FID evaluated with CCE1 and CCE2 with semiclassical hyperfine tensor components: absolute value of the coherence of a neutral $\mathrm{V}_{\mathrm{Si}}$ in 3C-SiC as a function of free evolution time. The external magnetic field is 200 G (a) and 500 G (b). The curves are averaged over 50 different baths.}
\label{fig:two graphs FID CCE1-2}
\end{figure}where 
$U_{\mathrm{FID}}(\tau)=e^{-i\mathcal{H}\tau}e^{-i\pi/2S_{x}}$ is the FID propagator, $\rho_{\mathrm{S}}(0)=\vert0\rangle\langle0\vert$ and the system's Hamiltonian is given by Eq. \ref{eq:hamtotpuredeph}. The real and imaginary parts of the coherence can be analytically calculated in the pure-dephasing approximation and CCE1 case, i.e. whenever we can write the Hamiltonian in the form of Eq. \ref{eq:hamtotpuredephcompact} and safely neglect the $\mathcal{H}_{n-n}$ interaction between nuclear spins inside the bath Hamiltonian in \ref{eq:hamtotpuredephcompactcond}. Then we substitute Eq. \ref{eq:hamtotpuredephcompact} in Eq. \ref{eq:totcoh} through $\rho_{\mathrm{tot}}(t)$, and the analytical expressions we obtain are the following~\cite{schweiger2001principles},
\begin{align}
\langle S_{x}\rangle_{\mathrm{FID}}(\tau)&=-\sin\left[\left(\omega_{1}-\omega_{0}\right)\tau\right]\mathrm{f}_{\mathrm{B}}(\tau), \label{eq:fidcohrepart} \\
\langle S_{y}\rangle_{\mathrm{FID}}(\tau)&=\cos\left[\left(\omega_{1}-\omega_{0}\right)\tau\right]\mathrm{f}_{\mathrm{B}}(\tau), \label{eq:fidcohimpart}
\end{align}
where
\begin{widetext}
\begin{equation}
\mathrm{f}_{\mathrm{B}}(\tau)=\prod_{i=1}^N\left[\cos\left(\frac{\omega_{I_{i}}\tau}{2}\right)\cos\left(\frac{\Omega_{I_{i}}\tau}{2}\right)+\sin\left(\frac{\omega_{I_{i}}\tau}{2}\right)\sin\left(\frac{\Omega_{I_{i}}\tau}{2}\right)\frac{\omega_{I_{i}}+A_{i}}{\Omega_{I_{i}}}\right] \label{eq:fidcohprod}
\end{equation}
\end{widetext}
is a factor depending on the nuclear spins, and
\begin{equation}
\Omega_{I_{i}}=\sqrt{\left(\omega_{I_{i}}+A_{i}\right)^2+B_{i}^2}. \label{eq:OmegaIi}    
\end{equation}
In Eq. \ref{eq:fidcohprod},
\begin{equation}
\omega_{I_{i}}=\gamma_{i}B \label{eq:nuclLarmfreq}    
\end{equation}
\noindent is the Larmor frequency of the $i$-th nuclear spin, where $B$ is the external magnetic field. \\
\hspace*{0.5cm}Our results on FID are displayed in Figs. \ref{fig:two graphs FID CCE1-2} and \ref{fig:two graphs FID CCE2 sm-abinit}. In Fig. \ref{fig:two graphs FID CCE1-2} we show a comparison between the absolute value of the qubit's coherence at the CCE1 and CCE2 levels of the theory, for two different external magnetic fields. The CCE1 curves exactly coincide with the analytical ones obtained as a graph of Eqs. \ref{eq:fidcohrepart} and \ref{eq:fidcohimpart}. Note that there is no interesting effect that is modeled in the passage from CCE1 to CCE2, and the two versions give pretty close results. In Fig. \ref{fig:two graphs FID CCE2 sm-abinit} we present the same curves at the CCE2 level, both with semiclassical and \textit{ab initio} hyperfine tensor components, for different external magnetic fields. As can be seen, the presence of even one single nuclear spin in the first shells of next-neighbors causes an appreciable change in the coherence, due to the difference in the hyperfine tensor components and therefore in $\Omega_{I_{i}}$ (remember that FID can be well-modeled already at CCE1, see Fig. \ref{fig:two graphs FID CCE1-2}). \\
\hspace*{0.5cm}In order to better understand our FID results and directly correlate differences in the position of the nuclear spins to the modification of the coherence modulation frequencies, we propose a manipulation of Eqs. \ref{eq:fidcohrepart} and \ref{eq:fidcohimpart}. In particular, by opportunely rewriting those equations we are able to explicitly obtain the coherence modulation frequencies. To do that we have to express the product of $N$ terms in Eq. \ref{eq:fidcohprod} as a sum of sinusoidal functions, by repeatedly applying the appropriate trigonometric formulas, so that the modulation frequencies are easily calculated via a Fourier transform. The new expressions can be obtained by exploiting induction considerations (see Appendix \ref{app:B}) and the imaginary part of the coherence, e.g., is given by
\begin{equation}
\langle S_{y}\rangle=\cos(\omega_{1}\tau)\Sigma_{N}(\tau), \label{eq:fidcohimpartsum}    
\end{equation}
\begin{figure}
\centering
\begin{subfigure}{0.4944\linewidth}
\caption{}
\label{fig:FID CCE2 sm-abinit 200G}
    \includegraphics[width=1.0\linewidth]{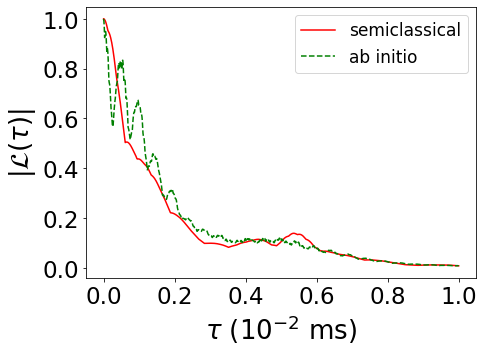}
\end{subfigure}
\hfill
\begin{subfigure}{0.4944\linewidth}
\caption{}
\label{fig:FID CCE2 sm-abinit 500G}
    \includegraphics[width=1.0\linewidth]{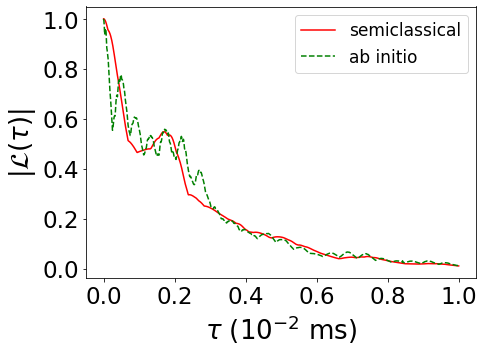}
\end{subfigure}
\caption{FID evaluated with CCE2 with semiclassical and \textit{ab initio} hyperfine tensor components: absolute value of the coherence of a neutral $\mathrm{V}_{\mathrm{Si}}$ in 3C-SiC as a function of free evolution time. The external magnetic field is 200 G (a) and 500 G (b). The curves are averaged over 50 different baths.}
\label{fig:two graphs FID CCE2 sm-abinit}
\end{figure}where
\begin{widetext}
\begin{equation} \label{eq:fidcohsum}
\begin{split}
\Sigma_{N}(\tau)&=\frac{1}{2^{2N-1}}\bigg\lbrace S_{I_{1}}\cdots S_{I_{N}}\Big[(+-+-\ldots+-)+\ldots+(+--+\ldots-+)\Big]\\
&+S_{I_{1}}\cdots D_{I_{N}}\Big[(+-+-\ldots++)+\ldots+(+--+\ldots--)\Big]+\ldots\\
&+S_{I_{1}}\cdots D_{I_{i}}\cdots D_{I_{N}}\Big[(+-\ldots++\ldots++)+\ldots+(+-\ldots--\ldots--)\Big]\\
&+\ldots+D_{I_{1}}\cdots D_{I_{N}}\Big[(++++\ldots++)+\ldots+(++--\ldots--)\Big]\bigg\rbrace.
\end{split}
\end{equation}
\end{widetext}\normalsize{In}
Eq. \ref{eq:fidcohsum} we have used the shorthand notation
\begin{equation}
\begin{split}
(+-\ldots+-)&\equiv\cos\big[(\omega_{I_{1}}/2-\Omega_{I_{1}}/2+\ldots \\
&+\omega_{I_{N}}/2-\Omega_{I_{N}}/2)\tau\big], \label{eq:fidshortnot}
\end{split}
\end{equation}
whereas
\begin{align}
S_{I_{i}}&\equiv1+\frac{\omega_{I_{i}}+A_{i}}{\Omega_{I_{i}}},\label{eq:SIi}\\
D_{I_{i}}&\equiv1-\frac{\omega_{I_{i}}+A_{i}}{\Omega_{I_{i}}}.\label{eq:DIi}\end{align}
Inside the curly brackets there are $2^N$ terms, each of which is multiplied by a sum of $2^{N-1}$ cosines inside the square brackets. Therefore, without counting the qubit through its level splitting $\omega_{1}$ (in which case the modulation frequencies would be doubled, see Appendix \ref{app:B}), the modulation frequencies are $2^N\times2^{N-1}=2^{2N-1}$. Eq. \ref{eq:fidcohsum} reduces to expected results in simple limiting conditions, i.e.
\begin{align}
&\Sigma_{N}\left(A=0,B=0\right)=1,\label{eq:fidcohsumtest1}\\
&\Sigma_{N}\left(\tau=0\right)=1.\label{eq:fidcohsumtest2}    
\end{align}
Eqs. \ref{eq:fidcohsumtest1} and \ref{eq:fidcohsumtest2}, coupled with Eq. \ref{eq:fidcohimpartsum}, give us the expected behavior of the coherence imaginary part when the qubit is isolated from the environment and at the beginning of the dynamics, respectively. Additional considerations regarding Eq. \ref{eq:fidcohsum} are reported in Appendix \ref{app:B}, in particular the numerical calculation of the modulation frequencies in a simple case. Finally, the modulation frequencies, containing information on how each nuclear spin in the bath affects the qubit during the dynamics, can be derived directly from the pure-dephasing Hamiltonian \ref{eq:hamtotpuredeph}, in the CCE1 case or whenever $\mathcal{H}_{n-n}=0$. In particular, they are obtained as linear combinations of our system's eigenenergies, as we demonstrate in Appendix \ref{app:C}. \\
\hspace*{0.5cm}At this point, the difference in the modulation frequencies in going from the semiclassical to the \textit{ab initio} curve in Fig. \ref{fig:two graphs FID CCE2 sm-abinit} is explained by considering the dependency of those frequencies, given in Eq. \ref{eq:fidcohsum}, on the hyperfine tensor components through $\Omega_{I_{i}}$. In particular, in the case where the bath is composed by a single $^{13}\mathrm{C}$ nucleus in the first neighbor shell, there are only two frequencies in the terms $(+-)$ and $(++)$. The first frequency doubles its value, from $13.1\hspace{0.07cm}\mathrm{MHz}$ to $29.4\hspace{0.07cm}\mathrm{MHz}$, by using the \textit{ab initio} calibration.

\subsection{\label{subsec:hahn-echo}Hahn-echo}

\begin{figure}
\centering
\begin{subfigure}{0.4944\linewidth}
\caption{}
\label{fig:Hahn echo graph 200G}
    \includegraphics[width=1.0\linewidth]{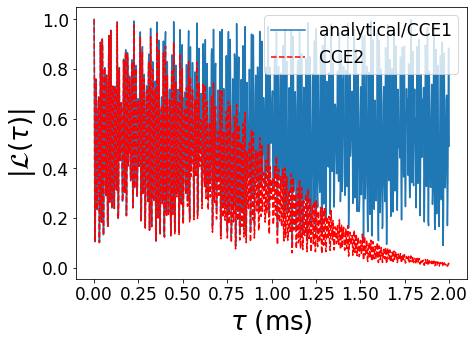}
\end{subfigure}
\hfill
\begin{subfigure}{0.4944\linewidth}
\caption{}
\label{fig:Hahn echo T2 graph 200G}
    \includegraphics[width=1.0\linewidth]{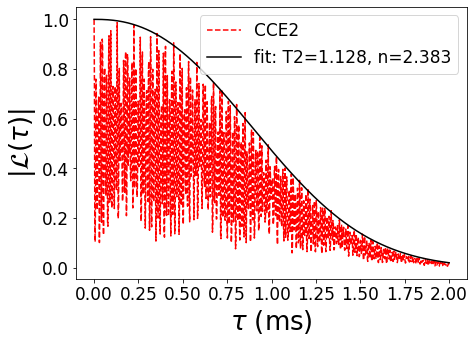}
\end{subfigure}
\caption{Hahn-echo evaluated with CCE1 and CCE2 with semiclassical hyperfine tensor components: absolute value of the coherence of a $\mathrm{V}_{\mathrm{Si}}^0$ in 3C-SiC as a function of time, for an external magnetic field of 200 G. The result is averaged over 50 different baths. (a): the blue (solid) curve is calculated analytically (Eq. \ref{eq:Hahnechocohimpart}) or at the CCE1 level and the red (dashed) one is calculated at the CCE2 level. The decoherence time is in the ms range. (b): fitting of the dashed curve of Fig. \ref{fig:Hahn echo graph 200G} with an exponential function $e^{-(t/T_{2})^n}$. Inset: the optimal values for the fitting parameters $T_{2}$, in ms, and $n$.}
\label{fig:two graphs Hahn-echo}
\end{figure}

\begin{figure}
\begin{center}
    \includegraphics[width=0.4944\textwidth]{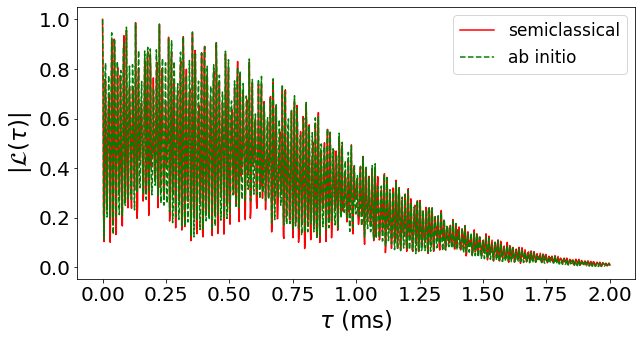}
    \caption{Hahn-echo evaluated with CCE2 with semiclassical (Eq. \ref{eq:smhyptens}) and \textit{ab initio} (Table \ref{tab:my_label}) hyperfine tensor components: absolute value of the coherence of a neutral $\mathrm{V}_{\mathrm{Si}}$ in 3C-SiC as a function of free evolution time, for an external magnetic field of 200 G. The result is averaged over 50 different baths.}
    \label{fig:comparison Hahn-echo sm-ab initio}
\end{center}
\end{figure}

In NMR/EPR systems environmental noise takes the form of magnetic field noise that results from the effect of accumulating disturbances from each nuclear spin-generated magnetic field (such static magnetic field inhomogeneity causes inhomogeneous broadening of the spectral lines~\cite{paladino20141}). To limit inhomogeneous broadening, we have applied the Hahn-echo sequence~\cite{okumura1992characterization}, an established control technique~\cite{schweiger2001principles,seo2016quantum} allowing to refocus the spin coherence and thus enlarge its decoherence time~\cite{fraval2005dynamic,seo2016quantum}, which is the main goal of this subsection. \\
\hspace*{0.5cm}In this regard, the most important part of the spin-echo sequence is an intermediate $\pi$ pulse applied to the qubit which allows to refocus the spin coherence resulting from the effect of static magnetic field inhomogeneities~\cite{slichter2013principles}. Consequently, the dynamics can be described in the following way:
\begin{equation}
\rho_{\mathrm{HE}}(\tau)=U_{\mathrm{HE}}(\tau)\rho_{\mathrm{S}}(0)U_{\mathrm{HE}}^\dagger(\tau), \label{eq:Hahnechoprocess}    
\end{equation}
where $U_{\mathrm{HE}}(\tau)=e^{-i\mathcal{H}\tau/2}e^{-i\pi S_{x}}e^{-i\mathcal{H}\tau/2}e^{-i\pi/2S_{x}}$ is the Hahn-echo propagator and $\rho_{\mathrm{S}}(0)$ is the same as for the FID case. Now, as a first order approximation, at the CCE1 level we can obtain analytical expressions for the coherence real and imaginary parts in the pure-dephasing approximation~\cite{seo2016quantum}, as in the FID case. Therefore, the qubit coherence components in \ref{eq:totcoh}, after the Hahn-echo sequence, can be written as~\cite{schweiger2001principles} 
\begin{align}
\langle S_{x}\rangle_{\mathrm{HE}}(\tau)&=0, \label{eq:Hahnechocohrepart} \\
\langle S_{y}\rangle_{\mathrm{HE}}(\tau)&=\prod_{i=1}^N\left[1-2k_{+1,0}^i\sin^2\left(\Omega_{I_{i}}\hspace{0.05cm}\frac{\tau}{4}\right)\sin^2\left(\omega_{I_{i}}\hspace{0.05cm}\frac{\tau}{4}\right)\right], \label{eq:Hahnechocohimpart}
\end{align}
where
\begin{equation}
k_{+1,0}^i=\frac{B_{i}^2}{\Omega_{I_{i}}^2}
\end{equation}
is the modulation depth parameter of the $i$-th nuclear spin between the $\vert0\rangle$ and $\vert+1\rangle$ qubit states. Eq. \ref{eq:Hahnechocohimpart} describes fast oscillations of the qubit coherence, or modulations (see Fig. \ref{fig:two graphs Hahn-echo}), known in the literature as Electron Spin Echo Envelope Modulation (ESEEM), which are due to single nuclear spin transitions~\cite{schweiger2001principles}. The real part of the coherence is zero also at $t=\tau$ because of the refocusing action of the central $\pi$ pulse. \\
\hspace*{0.5cm}Our results on the Hahn-echo extend the findings in Seo et al.~\cite{seo2016quantum} to a neutral $\mathrm{V}_{\mathrm{Si}}$ in 3C-SiC. Indeed the coherence decay is already obtained at the CCE2 level, as shown in Fig. \ref{fig:Hahn echo graph 200G} (this does not preclude the possibility of having further effects beyond CCE2). The figure shows the qubit coherence as a function of free evolution time, for an external magnetic field of 200 G. The solid curve is calculated at the CCE1 level, and exactly coincides with the analytical curve obtained as a graph of Eq. \ref{eq:Hahnechocohimpart}, as in the FID case. This should be the case since Eq. \ref{eq:Hahnechocohimpart} is obtained precisely by following the analytical counterpart of the numerical procedure behind the CCE1 approach, i.e. by neglecting $\mathcal{H}_{n-n}$ in \ref{eq:hamtotpuredephcompactcond} and thereby considering the coherence as a product of independent contributions coming from each nuclear spin. On the other hand, the dashed curve in Fig. \ref{fig:two graphs Hahn-echo} is calculated at the CCE2 level and presents the decay (note the difference with FID, for which CCE1 and CCE2 give similar results). Thus, interactions of the qubit with pairs of coupled nuclear spins within the bath cause a coherence decay that survives to the spin-echo protocol~\cite{seo2016quantum}. Furthermore, we demonstrate that the coherence decay of a $\mathrm{V}_{\mathrm{Si}}^0$ is in the ms range (see Fig. \ref{fig:Hahn echo T2 graph 200G}), whereas for FID it is in the 0.01 ms range (this difference is crucial in QT applications). The figure shows a fitting of the dashed curve of Fig. \ref{fig:Hahn echo graph 200G} with a stretched exponential function. The fitting curve's parameters are the Hahn-echo decoherence time and the stretching factor, whose optimal values are $T_{2}=1.13\hspace{0.1cm}\mathrm{ms}$ and $n=2.38$, respectively. Similar values of the decay parameters ($T_{2}=1.16\hspace{0.1cm}\mathrm{ms}$ and $n=2.05$) have been derived by means of a CCE2 calculation of the Hahn-echo protocol generalized to the $S=3/2$ spin state of $\mathrm{V}_{\mathrm{Si}}^{-1}$ (see Appendix \ref{app:D}). Thus, the decoherence times of both $\mathrm{V}_{\mathrm{Si}}^0$ and  $\mathrm{V}_{\mathrm{Si}}^{-1}$ in 3C-SiC have the same order of magnitude as the ones associated to NV centers in diamond~\cite{bauch2020decoherence} and divacancies in 4H-SiC~\cite{seo2016quantum}. Moreover, although the optical initialization and readout of spin states for our defect is not remotely close to the level of sophistication for, e.g., NV centers in SiC~\cite{von2016nv}, we hope that our work will encourage experimental analyses in this direction. Due to the presence of a stretching factor, we demonstrate that also for a $\mathrm{V}_{\mathrm{Si}}$ in 3C-SiC the decay of the coherence envelope is not exponential, which is a typical behavior for qubits in NMR/EPR and in general in the solid state. As a matter of fact, this also happens for superconducting qubits, which are usually subjected to $1/f$-type noise from the environment~\cite{paladino20141}. \\
\hspace*{0.5cm}Then, we have used the \textit{ab initio} calculated values of the hyperfine tensor components listed in Table \ref{tab:my_label} in our CCE code. The comparison of the resulting coherence curve with the semiclassical one, for an external applied magnetic field of 200 G, is shown in Fig. \ref{fig:comparison Hahn-echo sm-ab initio}. The main difference is in the modulation effect, whereas the decay, and hence the decoherence time, is almost unchanged. Again, doing the comparison with FID we see that spin-echo protocols are more robust against the hyperfine tensor components change due to the \textit{ab initio} calibration. This is due to the refocusing $\pi$ pulse that lifts the dependence on one-body interactions, which are more affected by the \textit{ab initio} calibration. This is in turn due to there being way more one-body interactions where the electron-nuclear spin distance is such that the \textit{ab initio} value is used as opposed to two-body interactions where \textit{both} nuclear spins are close enough to require the \textit{ab initio} calibration. This behavior is understood by looking at Eq. \ref{eq:Hahnechocohimpart}, for which a similar reasoning used in the passage from Eq. \ref{eq:fidcohimpart} to Eq. \ref{eq:fidcohimpartsum} can be applied to analytically calculate the modulation frequencies. Those frequencies depend both on the single nuclear spin Larmor frequencies and the hyperfine tensor components through $\Omega_{I_{i}}$. Therefore, if any of the $50$ random baths in a given simulation happens to have a nuclear spin in the first or second neighbor shell, the hyperfine tensor components entering Eq. \ref{eq:Hahnechocohimpart}, and thereby the modulations of the coherence, will be modified. We find the change of the modulation frequencies by using the \textit{ab initio} calibration to be of the same order of magnitude as in the FID case (see the last paragraph of the previous subsection). As for the decoherence effect, which is at least caused by two-body interactions between nuclear spins (it appears at least at the CCE2 level), the probability of having two nuclear spins in the first and second neighbor shell is less than the probability of having just one, thus conditioning less the coherence decay. This is a consequence of the chosen numerical random bath-generating procedure.

\section{\label{sec:discussion}Discussion}

In this section we summarize the findings of our work. We have employed \textit{ab initio} methods to calculate the formation energy and the magnetic parameters of a neutral and charged $\mathrm{V}_{\mathrm{Si}}$ in 3C-SiC, indicating the stability of the less studied neutral charge state of the $\mathrm{V}_{\mathrm{Si}}$ for energies near to the valence band maximum, particularly in the presence of the monopole correction term, whose effect is to widen the electrochemical potential range over which this stability holds. We have performed magnetic calculations for the $\mathrm{V}_{\mathrm{Si}}^0$  and $\mathrm{V}_{\mathrm{Si}}^{-1}$ in 3C-SiC and for the $\mathrm{V}_{\mathrm{Si}}^0$ in 4H-SiC in order to compare the results on the hyperfine tensor components in several cases, finding a significant deviation with respect to the semiclassical values of the model parameters obtained with Eq. \ref{eq:smhyptens}. We have studied both Hahn spin-echo and FID as protocols applied to our $\mathrm{V}_{\mathrm{Si}}^0$ qubit, finding the equations (Eqs. \ref{eq:fidcohrepart} and \ref{eq:fidcohimpart}) that describe FID when the nuclear spins are non-interacting with each other. By appropriately rewriting these equations, we have been able to analytically calculate, at the CCE1 level, the FID modulation frequencies and directly associate them to the Hamiltonian eigenvalues, and hence to the system's magnetic parameters. Furthermore, by applying the CCE theory, we have shed some light on the ESEEM phenomenon and the decoherence of the qubit, after applying the FID and Hahn-echo processes. We calculated the Hahn-echo decoherence time associated to a $\mathrm{V}_{\mathrm{Si}}^0$ and $\mathrm{V}_{\mathrm{Si}}^{-1}$  in 3C-SiC to be in the ms range, thus gaining at least two orders of magnitude with respect to FID. We have also evaluated the non-exponential character of the coherence decay, which is typical for qubits in solid state devices. Finally, for the spin-echo we have demonstrated that modeling the three-dimensional distribution of the defect's spin density in our CCE simulations has an effect on the coherence modulations (ESEEM), but not as much on the decoherence effect, which is important for QT applications. For the FID process the effect is instead more pronounced. This is due to the FID process being dominated by one-body interactions between the qubit and the nuclear spins in the bath, and to these interactions being more affected by the \textit{ab initio} calibration than the two-body ones between nuclear spins. \\
%\hspace*{0.5cm}{\color{red} As a last point, it is useful to discuss how our analysis would be modified by a change in the structural model describing the defect, e.g. a different symmetry, since this issue is not yet settled in the literature. A $C_{3v}$ symmetry would give rise to an axial ZFS tensor, so that our approach would not be appreciably modified due to the fact that $E\ll D$ (Section \ref{subsec:hyp_int_zfs})}. \\
%\hspace*{0.5cm}Part of our findings is available as we point out in the data availability statement.

\section{\label{sec:methods}Methods}

\subsection{\label{subsec:ab_initio_calc}\textit{Ab initio} calculations}

In this study we have used the Quantum Espresso (QE) code~\cite{giannozzi2009quantum} for calculating total energies and magnetic properties of the $\mathrm{V}_{\mathrm{Si}}$ in 3C-SiC and 4H-SiC. For the cubic polytype we considered a $7\times7\times7$ 3C-SiC supercell starting from a primitive fcc (face-centered cubic) unit cell, containing 686 atoms, for the calculation of the formation energy~\cite{xi2016ab}, whereas a $6\times6\times6$ supercell, containing 432 atoms, was employed for the evaluation of the EPR-related parameters like the hyperfine and ZFS tensor components. Hexagonal systems were studied in $5\times5\times2$ supercells. We used the Perdew-Burke-Ezernhof implementation~\cite{perdew1996generalized} of the generalized gradient approximation for the description of the exchange-correlation functional. Ultrasoft pseudopotentials~\cite{vanderbilt1990soft} were used for standard ground-state properties, whereas hyperfine interactions and ZFS tensors were computed with norm-conserving pseudopotentials~\cite{hamann1979norm}, as the latter showed a better agreement with respective experimental results~\cite{ivady2017identification}. The formation energy was evaluated for the $\mathrm{V}_{\mathrm{Si}}$ in various charged configurations considering a non-collinear scheme for the magnetization. EPR calculations were instead performed by using a collinear magnetization along the $[001]$ lattice direction, in order to better comply with usual experimental setups, where the external magnetic field is applied along the growth direction (which coincides with the $[001]$ crystal direction for most 3C-SiC growths~\cite{fisicaro2020genesis}). Convergence was achieved with an asymmetric $3\times3\times3$ k-point grid~\cite{monkhorst1976special} having an offset with respect to the $\Gamma$ point. Upon completion of the DFT calculations, the QE wave functions were used as an input in the QE Gauge-Including Projector Augmented-Wave (QE-GIPAW) code~\cite{pickard2001all,varini2013enhancement}, to calculate the hyperfine tensor components describing the $\mathrm{V}_{\mathrm{Si}}$-nuclear spins interaction. The spin density in the vicinity of the nucleus was evaluated by applying a transformation that reconstructs the all-electron wave functions in the core regions~\cite{blochl1994projector}, followed by a first-order perturbation theory approach in which the perturbing potential is a functional of charge and spin densities (considering only its exchange part). This procedure, implemented in the GIPAW code~\cite{bahramy2007pseudopotential} is particularly important for the calculation of the Fermi-contact term of the hyperfine tensor. In addition, QE wave functions were used to calculate the ZFS tensor components with the aid of the PyZFS code~\cite{ma2020pyzfs}. Finally, in order to better understand if the stacking sequence of the SiC polytype has important implications in the magnetic properties of the defect, hyperfine tensor calculations were also performed for a neutral $\mathrm{V}_{\mathrm{Si}}$ in hexagonal 4H-SiC (for both k and h sites). \\

\begin{figure}
\begin{center}
\tikzset{every picture/.style={line width=0.75pt}} %set default line width to 0.75pt        

\begin{tikzpicture}[x=0.75pt,y=0.75pt,yscale=-0.7944,xscale=0.7944]

%Flowchart: Connector [id:dp7559353510493512] 
\draw   (267,121) .. controls (267,72.4) and (310.74,33) .. (364.7,33) .. controls (418.66,33) and (462.4,72.4) .. (462.4,121) .. controls (462.4,169.6) and (418.66,209) .. (364.7,209) .. controls (310.74,209) and (267,169.6) .. (267,121) -- cycle ;
%Up Arrow [id:dp5356917261630194] 
\draw  [color={rgb, 255:red, 74; green, 144; blue, 226 }  ,draw opacity=1 ][fill={rgb, 255:red, 74; green, 144; blue, 226 }  ,fill opacity=1 ] (292.02,113.01) -- (295.81,97) -- (299.42,112.99) -- (297.57,113) -- (297.44,137) -- (293.74,137) -- (293.87,113) -- cycle ;
%Down Arrow [id:dp05047896032466115] 
\draw  [color={rgb, 255:red, 74; green, 144; blue, 226 }  ,draw opacity=1 ][fill={rgb, 255:red, 74; green, 144; blue, 226 }  ,fill opacity=1 ] (428,124) -- (430.1,124) -- (430.1,100) -- (434.3,100) -- (434.3,124) -- (436.4,124) -- (432.2,140) -- cycle ;
%Down Arrow [id:dp39074349646354545] 
\draw  [color={rgb, 255:red, 208; green, 2; blue, 27 }  ,draw opacity=1 ][fill={rgb, 255:red, 208; green, 2; blue, 27 }  ,fill opacity=1 ] (359,176) -- (361.1,176) -- (361.1,152) -- (365.3,152) -- (365.3,176) -- (367.4,176) -- (363.2,192) -- cycle ;
%Up Arrow [id:dp9154647397812439] 
\draw  [color={rgb, 255:red, 208; green, 2; blue, 27 }  ,draw opacity=1 ][fill={rgb, 255:red, 208; green, 2; blue, 27 }  ,fill opacity=1 ] (357,61) -- (361.2,45) -- (365.4,61) -- (363.3,61) -- (363.3,85) -- (359.1,85) -- (359.1,61) -- cycle ;
%Straight Lines [id:da12387150516138301] 
\draw  [dash pattern={on 4.5pt off 4.5pt}]  (308.98,96.47) -- (347.82,73.53) ;
\draw [shift={(350.4,72)}, rotate = 149.42] [fill={rgb, 255:red, 0; green, 0; blue, 0 }  ][line width=0.08]  [draw opacity=0] (10.72,-5.15) -- (0,0) -- (10.72,5.15) -- (7.12,0) -- cycle    ;
\draw [shift={(306.4,98)}, rotate = 329.42] [fill={rgb, 255:red, 0; green, 0; blue, 0 }  ][line width=0.08]  [draw opacity=0] (10.72,-5.15) -- (0,0) -- (10.72,5.15) -- (7.12,0) -- cycle    ;
%Straight Lines [id:da6629249512839472] 
\draw  [dash pattern={on 4.5pt off 4.5pt}]  (308.77,146.83) -- (348.03,177.17) ;
\draw [shift={(350.4,179)}, rotate = 217.69] [fill={rgb, 255:red, 0; green, 0; blue, 0 }  ][line width=0.08]  [draw opacity=0] (10.72,-5.15) -- (0,0) -- (10.72,5.15) -- (7.12,0) -- cycle    ;
\draw [shift={(306.4,145)}, rotate = 37.69] [fill={rgb, 255:red, 0; green, 0; blue, 0 }  ][line width=0.08]  [draw opacity=0] (10.72,-5.15) -- (0,0) -- (10.72,5.15) -- (7.12,0) -- cycle    ;
%Straight Lines [id:da7247959713377454] 
\draw  [dash pattern={on 4.5pt off 4.5pt}]  (315.4,118.97) -- (415.4,118.03) ;
\draw [shift={(418.4,118)}, rotate = 179.46] [fill={rgb, 255:red, 0; green, 0; blue, 0 }  ][line width=0.08]  [draw opacity=0] (10.72,-5.15) -- (0,0) -- (10.72,5.15) -- (7.12,0) -- cycle    ;
\draw [shift={(312.4,119)}, rotate = 359.46] [fill={rgb, 255:red, 0; green, 0; blue, 0 }  ][line width=0.08]  [draw opacity=0] (10.72,-5.15) -- (0,0) -- (10.72,5.15) -- (7.12,0) -- cycle    ;
%Straight Lines [id:da877225361948524] 
\draw  [dash pattern={on 4.5pt off 4.5pt}]  (362.02,95) -- (362.38,141) ;
\draw [shift={(362.4,144)}, rotate = 269.56] [fill={rgb, 255:red, 0; green, 0; blue, 0 }  ][line width=0.08]  [draw opacity=0] (10.72,-5.15) -- (0,0) -- (10.72,5.15) -- (7.12,0) -- cycle    ;
\draw [shift={(362,92)}, rotate = 89.56] [fill={rgb, 255:red, 0; green, 0; blue, 0 }  ][line width=0.08]  [draw opacity=0] (10.72,-5.15) -- (0,0) -- (10.72,5.15) -- (7.12,0) -- cycle    ;
%Straight Lines [id:da229884957706058] 
\draw  [dash pattern={on 4.5pt off 4.5pt}]  (377.53,69.6) -- (419.87,96.4) ;
\draw [shift={(422.4,98)}, rotate = 212.33] [fill={rgb, 255:red, 0; green, 0; blue, 0 }  ][line width=0.08]  [draw opacity=0] (10.72,-5.15) -- (0,0) -- (10.72,5.15) -- (7.12,0) -- cycle    ;
\draw [shift={(375,68)}, rotate = 32.33] [fill={rgb, 255:red, 0; green, 0; blue, 0 }  ][line width=0.08]  [draw opacity=0] (10.72,-5.15) -- (0,0) -- (10.72,5.15) -- (7.12,0) -- cycle    ;
%Straight Lines [id:da5445579567524881] 
\draw  [dash pattern={on 4.5pt off 4.5pt}]  (380.82,172.23) -- (420.98,142.77) ;
\draw [shift={(423.4,141)}, rotate = 143.75] [fill={rgb, 255:red, 0; green, 0; blue, 0 }  ][line width=0.08]  [draw opacity=0] (10.72,-5.15) -- (0,0) -- (10.72,5.15) -- (7.12,0) -- cycle    ;
\draw [shift={(378.4,174)}, rotate = 323.75] [fill={rgb, 255:red, 0; green, 0; blue, 0 }  ][line width=0.08]  [draw opacity=0] (10.72,-5.15) -- (0,0) -- (10.72,5.15) -- (7.12,0) -- cycle    ;
%Up Arrow [id:dp5517792738405325] 
\draw  [color={rgb, 255:red, 208; green, 2; blue, 27 }  ,draw opacity=1 ][fill={rgb, 255:red, 208; green, 2; blue, 27 }  ,fill opacity=1 ] (67,226) -- (71.2,210) -- (75.4,226) -- (73.3,226) -- (73.3,250) -- (69.1,250) -- (69.1,226) -- cycle ;
%Up Arrow [id:dp9951069842385563] 
\draw  [color={rgb, 255:red, 74; green, 144; blue, 226 }  ,draw opacity=1 ][fill={rgb, 255:red, 74; green, 144; blue, 226 }  ,fill opacity=1 ] (166,228) -- (169.7,212) -- (173.4,228) -- (171.55,228) -- (171.55,252) -- (167.85,252) -- (167.85,228) -- cycle ;
%Straight Lines [id:da9755237905430099] 
\draw  [dash pattern={on 4.5pt off 4.5pt}]  (238.23,218.01) -- (268.17,244.99) ;
\draw [shift={(270.4,247)}, rotate = 222.02] [fill={rgb, 255:red, 0; green, 0; blue, 0 }  ][line width=0.08]  [draw opacity=0] (10.72,-5.15) -- (0,0) -- (10.72,5.15) -- (7.12,0) -- cycle    ;
\draw [shift={(236,216)}, rotate = 42.02] [fill={rgb, 255:red, 0; green, 0; blue, 0 }  ][line width=0.08]  [draw opacity=0] (10.72,-5.15) -- (0,0) -- (10.72,5.15) -- (7.12,0) -- cycle    ;
%Flowchart: Connector [id:dp7255077214225572] 
\draw   (51,121) .. controls (51,72.4) and (94.74,33) .. (148.7,33) .. controls (202.66,33) and (246.4,72.4) .. (246.4,121) .. controls (246.4,169.6) and (202.66,209) .. (148.7,209) .. controls (94.74,209) and (51,169.6) .. (51,121) -- cycle ;
%Up Arrow [id:dp4222115037650367] 
\draw  [color={rgb, 255:red, 74; green, 144; blue, 226 }  ,draw opacity=1 ][fill={rgb, 255:red, 74; green, 144; blue, 226 }  ,fill opacity=1 ] (82.02,113.01) -- (85.81,97) -- (89.42,112.99) -- (87.57,113) -- (87.44,137) -- (83.74,137) -- (83.87,113) -- cycle ;
%Down Arrow [id:dp5010458631008008] 
\draw  [color={rgb, 255:red, 74; green, 144; blue, 226 }  ,draw opacity=1 ][fill={rgb, 255:red, 74; green, 144; blue, 226 }  ,fill opacity=1 ] (205,122) -- (207.1,122) -- (207.1,98) -- (211.3,98) -- (211.3,122) -- (213.4,122) -- (209.2,138) -- cycle ;
%Down Arrow [id:dp2076018869258045] 
\draw  [color={rgb, 255:red, 208; green, 2; blue, 27 }  ,draw opacity=1 ][fill={rgb, 255:red, 208; green, 2; blue, 27 }  ,fill opacity=1 ] (143,174) -- (145.1,174) -- (145.1,150) -- (149.3,150) -- (149.3,174) -- (151.4,174) -- (147.2,190) -- cycle ;
%Up Arrow [id:dp9479803181300437] 
\draw  [color={rgb, 255:red, 208; green, 2; blue, 27 }  ,draw opacity=1 ][fill={rgb, 255:red, 208; green, 2; blue, 27 }  ,fill opacity=1 ] (141,60) -- (145.2,44) -- (149.4,60) -- (147.3,60) -- (147.3,84) -- (143.1,84) -- (143.1,60) -- cycle ;
%Shape: Circle [id:dp9960277137426106] 
\draw  [dash pattern={on 4.5pt off 4.5pt}] (120.2,69) .. controls (120.2,55.19) and (131.39,44) .. (145.2,44) .. controls (159.01,44) and (170.2,55.19) .. (170.2,69) .. controls (170.2,82.81) and (159.01,94) .. (145.2,94) .. controls (131.39,94) and (120.2,82.81) .. (120.2,69) -- cycle ;
%Shape: Circle [id:dp638965766388653] 
\draw  [dash pattern={on 4.5pt off 4.5pt}] (62.57,113) .. controls (62.57,99.19) and (73.77,88) .. (87.57,88) .. controls (101.38,88) and (112.57,99.19) .. (112.57,113) .. controls (112.57,126.8) and (101.38,138) .. (87.57,138) .. controls (73.77,138) and (62.57,126.8) .. (62.57,113) -- cycle ;
%Shape: Circle [id:dp08791233012294364] 
\draw  [dash pattern={on 4.5pt off 4.5pt}] (184.2,113) .. controls (184.2,99.19) and (195.39,88) .. (209.2,88) .. controls (223.01,88) and (234.2,99.19) .. (234.2,113) .. controls (234.2,126.81) and (223.01,138) .. (209.2,138) .. controls (195.39,138) and (184.2,126.81) .. (184.2,113) -- cycle ;
%Shape: Circle [id:dp8210261381407709] 
\draw  [dash pattern={on 4.5pt off 4.5pt}] (122.2,165) .. controls (122.2,151.19) and (133.39,140) .. (147.2,140) .. controls (161.01,140) and (172.2,151.19) .. (172.2,165) .. controls (172.2,178.81) and (161.01,190) .. (147.2,190) .. controls (133.39,190) and (122.2,178.81) .. (122.2,165) -- cycle ;
%Shape: Ellipse [id:dp8567775168267289] 
\draw  [dash pattern={on 4.5pt off 4.5pt}] (286.39,90) .. controls (302.97,63.49) and (335.21,42) .. (358.41,42) .. controls (381.6,42) and (386.98,63.49) .. (370.41,90) .. controls (353.83,116.51) and (321.59,138) .. (298.39,138) .. controls (275.2,138) and (269.82,116.51) .. (286.39,90) -- cycle ;
%Straight Lines [id:da06726294505087971] 
\draw  [dash pattern={on 4.5pt off 4.5pt}]  (368,217) -- (399.4,249) ;

% Text Node
\draw (86,220) node [anchor=north west][inner sep=0.75pt]    {$^{29}\mathrm{Si}$};
% Text Node
\draw (183,220) node [anchor=north west][inner sep=0.75pt]    {$^{13}\mathrm{C}$};
% Text Node
\draw (273,222) node [anchor=north west][inner sep=0.75pt]   [align=left] {\textbf{Interaction}};
% Text Node
\draw (61,18) node [anchor=north west][inner sep=0.75pt]   [align=left] {(a)};
% Text Node
\draw (275,19) node [anchor=north west][inner sep=0.75pt]   [align=left] {(b)};
% Text Node
\draw (399,222) node [anchor=north west][inner sep=0.75pt]   [align=left] {\textbf{Cluster}};

\end{tikzpicture}
\caption{Functioning scheme of the CCE theory, for a generic bath in 3C-SiC containing two $^{29}\mathrm{Si}$ nuclear spins (in red) and two $^{13}\mathrm{C}$ nuclear spins (in blue). The spins are in general all interacting with each other via the last term of Eq. \ref{eq:hamtotpuredeph} and are represented in the up or down states with respect to the magnetic field axis. (a): CCE1 approximation in which clusters contain a single nuclear spin. (b): CCE2 approximation containing also two-dimensional clusters (for simplicity only one is shown).}
\label{fig:CCE scheme}
\end{center}
\end{figure}

\subsection{\label{subsec:cce_theory}CCE theory}

In 3C-SiC samples there are thousands of paramagnetic nuclear spins, each of which exerts an influence on our qubit. In order to deal with nuclear spins in large baths, various theories have been introduced in the literature. Among them, CCE theory is particularly useful to calculate the qubit's coherence, which is our objective. CCE theory has been developed in reference~\cite{yang2008quantum} and is perfectly suited for qubits experiencing random interactions within a bath of finite size. As a matter of fact, whenever there are few nuclear spins in the bath, the qubit may not complete its decoherence within the nuclear spin flip-flop time and higher-order cluster correlations (cfr. Fig. \ref{fig:CCE scheme}) could be necessary to model the dynamics. In this case, among the various theories developed such as the density matrix Cluster Expansion (CE)~\cite{witzel2006quantum}, the pair-correlation approximation~\cite{yao2006theory} and the Linked-Cluster Expansion (LCE)~\cite{saikin2007single}, only the CCE converges to the exact coherent dynamics of clusters containing multiple spins. In particular, a cluster in this context is defined as a group of fully interacting nuclear spins. CCE theory owes its high convergence property to the fact that it is a bridge between the LCE and CE approaches. One is not required to evaluate Feynman diagrams and is simultaneously free from the large-bath restriction of the CE. However, typically CCE theory does not converge whenever its $N$-th truncation, or CCE$N$ (see below), is not sufficient to model the dynamics. In this case a small term in the recursive expansion in the denominator of Eq. \ref{eq:Cclustcohcce} below is not balanced by a similar next-order term in the numerator and the final result blows up, thus lying outside of the expected range for coherence. The coherent dynamics of finite clusters of nuclear spins in the bath is of special interest in systems with random couplings between the qubit and bath. Interesting examples are nitrogen-vacancy (NV) centers in diamond and $\mathrm{V}_{\mathrm{Si}}$ in SiC, which are magnetically coupled to randomly located nuclear spins in the vicinity. For such systems, the analysis in reference~\cite{yao2006theory} taking into account only pair-correlations is not sufficient, e.g., to describe free evolution, which is governed by singular interactions between the qubit and the nuclear spins. CCE theory has the advantage of being in principle exact, while simultaneously being of great practical utility as an approximation scheme whenever many-body correlations within the bath are not relevant \textit{and} being more flexible than pair-correlation approaches when higher-order correlations are needed. \\
\hspace*{0.5cm}In the CCE theory the spin baths is considered in a thermal equilibrium state at $t=0$~\cite{yang2008quantum}.
At the typical experimental temperatures of 
%In typical EPR experiments the temperature of the examined sample is of 
$\sim10\hspace{0.1cm}\mathrm{K}$~\cite{seo2016quantum,schweiger2001principles}, it is reasonable to consider a completely randomized bath. In our CCE code we use the mixed state
\begin{equation}
\rho_{\mathrm{B}}(0)=\bigotimes_{i=1}^N\frac{I_{i}}{2}, \label{eq:initbathstate}    
\end{equation}where 
$I_{i}$ is the $i$-th nuclear spin identity operator. 
%{\color{red} While temperature is not explicitly present in the CCE equations, this is the indirect way to account for it}. 
Furthermore, in order to obtain a correct statistical sampling of the random bath-generating procedure, we calculate the qubit coherence as an average of the coherences associated to $\mathcal{N}$ different random baths, or $\mathcal{N}$ different realizations of the numerical procedure. Hence, in calculating the qubit coherence we confirmed that the converged value~\cite{seo2016quantum} for $\mathcal{N}$ in our simulations is $\mathcal{N}=50$ (see Appendix \ref{app:A}). In the generation of the random baths, other numerical parameters whose convergence is necessary are the radius of the spherical bath, $R_{\mathrm{bath}}$, and the distance between nuclear spins beyond which they are no longer interacting, or nuclear spin connectivity, $r_{\mathrm{dipole}}$. The converged values for these parameters (Appendix \ref{app:A}) are found to be $R_{\mathrm{bath}}=5\hspace{0.1cm}\mathrm{nm}$ and $r_{\mathrm{dipole}}=0.8\hspace{0.1cm}\mathrm{nm}$, as in reference~\cite{seo2016quantum}. \\
\hspace*{0.5cm}Once all of this is taken care of, we have implemented CCE theory through the equations~\cite{onizhuk2021probing}
\begin{align}
\rho_{C}(t)&=U_{C}\rho_{C}(0)U_{C}^\dagger, \label{eq:Cclustdyn} \\
\Tilde{\mathcal{L}}_{\lbrace C\rbrace}&=\frac{\mathcal{L}_{\lbrace C\rbrace}(t)}{\prod_{C'}\Tilde{\mathcal{L}}_{\lbrace C'\subset C\rbrace}}, \label{eq:Cclustcohcce} \\
\mathcal{L}(t)&=\Tilde{\mathcal{L}}_{\lbrace0\rbrace}\prod_{i}\Tilde{\mathcal{L}}_{\lbrace i\rbrace}\prod_{i,j}\Tilde{\mathcal{L}}_{\lbrace ij\rbrace}\cdots. \label{eq:totcohcce}
\end{align}
Eq. \ref{eq:Cclustdyn} describes the dynamics of the density matrix of the qubit interacting with a given cluster of nuclear spins C, the Hamiltonian in the time evolution operator $U_{C}$ being given by Eq. \ref{eq:hamtotpuredeph} restricted to the sole presence of the nuclear spins within cluster C. Eq. \ref{eq:Cclustdyn} enters in Eq. \ref{eq:Cclustcohcce} through $\mathcal{L}_{\lbrace C\rbrace}(t)$, which can be written as
\begin{equation}
\mathcal{L}_{\lbrace C\rbrace}(t)=\frac{\mathrm{tr}\left\lbrace\rho_{C}(t)S_{+}\right\rbrace}{\mathrm{tr}\left\lbrace\rho_{C}(0)S_{+}\right\rbrace}. \label{eq:clustercoh}
\end{equation}
Eq. \ref{eq:Cclustcohcce} describes the contribution of cluster C to the coherence. Since the clusters are uncorrelated, the coherence in Eq. \ref{eq:totcohcce} is defined as the product of each cluster's contribution. As a consequence, we can stop the expansion in Eq. \ref{eq:totcohcce} at a given order of approximation of the theory, which is represented by the number of nuclear spins within the largest clusters we choose to divide the bath in. Therefore, CCE$N$ is the implementation of CCE theory where the biggest clusters we consider contain $N$ different nuclear spins. In this paper we focus exclusively on the CCE1 and CCE2 approaches, with clusters containing single and interacting pairs of nuclear spins, respectively (see Fig. \ref{fig:CCE scheme} for the functioning scheme of the CCE1 and CCE2 approaches).

\begin{acknowledgments}
The work has been partially funded by the Italian Ministry University and Research (MUR) in the framework of project PNRR Partenariato PE4 NQSTI Quantum, Grant No. PE0000023. T.F. acknowledges a fruitful conversation and exchange of emails with Mykyta Onizhuk from the group of Giulia Galli and with Dr. Davide Ceresoli from the SCITEC institute in Milan.
\end{acknowledgments}

\appendix

\section{\label{app:A}Convergence and choice of parameters}

\begin{figure}[b]
\centering
\begin{subfigure}{0.4944\linewidth}
\caption{}
\label{fig:var D FID 500G}
    \includegraphics[width=1.0\linewidth]{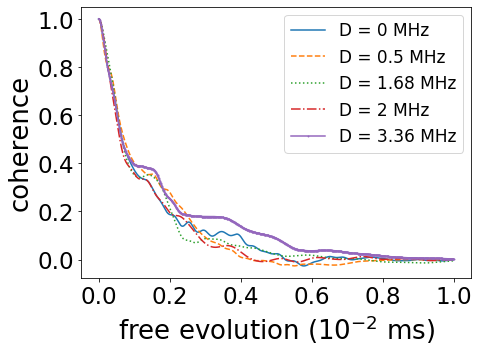}
\end{subfigure}
\hfill
\begin{subfigure}{0.4944\linewidth}
\caption{}
\label{fig:var D and E FID 500G}
    \includegraphics[width=1.0\linewidth]{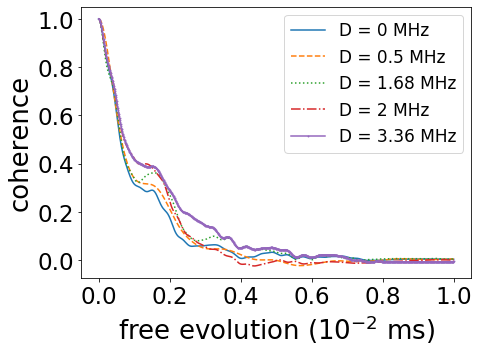}
\end{subfigure}
\caption{CCE2 absolute value of the coherence of a neutral $\mathrm{V}_{\mathrm{Si}}$ in 3C-SiC as a function of free evolution time, with different values for the axial component $D$ of the ZFS tensor, for an external magnetic field of 500 G. The dynamics implemented is a free evolution, or FID process, and the results are averaged over 50 different baths. (a): the transversal component $E$ of the ZFS tensor is absent. (b): the transversal component $E$ of the ZFS tensor is present ($E=-0.41\hspace{0.1cm}\mathrm{MHz}$). The curves are obtained with the PyCCE code~\cite{onizhuk2021pycce}.}
\label{fig:two graphs var D FID}
\end{figure}

\begin{figure}
\centering
\begin{subfigure}{0.4944\linewidth}
\caption{}
\label{fig:var R FID 340G}
    \includegraphics[width=0.8579\linewidth]{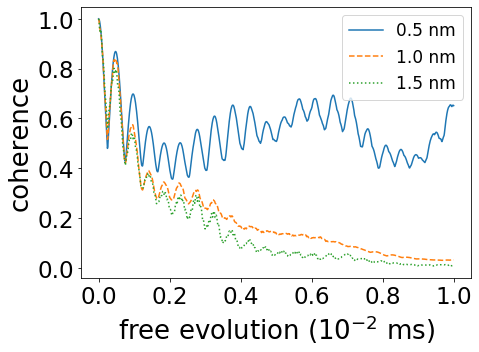}
\end{subfigure}
\hfill
\begin{subfigure}{0.4944\linewidth}
\caption{}
\label{fig:var R FID 500G}
    \includegraphics[width=0.8579\linewidth]{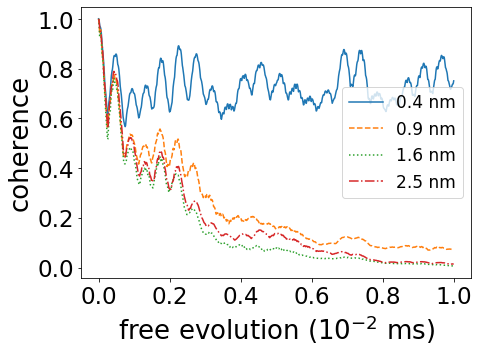}
\end{subfigure}
\caption{CCE1 absolute value of the coherence of a neutral $\mathrm{V}_{\mathrm{Si}}$ in 3C-SiC as a function of free evolution time, with different values for the radius of the spherical bath $R_{\mathrm{bath}}$, for an external magnetic field of 340 G (a) and 500 G (b). The dynamics implemented is a free evolution, or FID process, and the results are averaged over 50 different baths.}
\label{fig:two graphs var R FID}
\end{figure}

In this section we present results on the convergence of the parameters of our simulations. Fig. \ref{fig:two graphs var D FID}(a) shows the dependence of the coherence on the axial component $D$ in a FID process when the transversal component $E$ of the ZFS tensor is absent. These results are obtained with the PyCCE code~\cite{onizhuk2021pycce}. Fig. \ref{fig:two graphs var D FID}(b) shows that the presence of $E=-0.41\hspace{0.1cm}\mathrm{MHz}$ value for the transversal component only marginally changes the result. Hence, our choice of neglecting $E$ is justified and further validates the \textit{ab initio} findings. In Fig. \ref{fig:two graphs var R FID} we study the dependence of the coherence on the radius of the spherical bath $R_{\mathrm{bath}}$ at the CCE1 level, which is sufficient for analyzing the FID. The figure shows that already for a bath with a radius of dimension $R_{\mathrm{bath}}=2.5\hspace{0.1cm}\mathrm{nm}$ the result is almost completely converged, justifying our choice of $R_{\mathrm{bath}}=5\hspace{0.1cm}\mathrm{nm}$. Fig. \ref{fig:two graphs var nbaths FID} shows the dependence of the coherence on the number of realizations of the bath $\mathcal{N}$, again at the CCE1 level and for a reduced bath of $R_{\mathrm{bath}}=4\hspace{0.1cm}\mathrm{nm}$. Note that already a mean over 50 different baths is sufficient to obtain a reasonably well-converged coherence curve, which is why we choose $\mathcal{N}=50$ in the main body of the paper. \\
\hspace*{0.5cm}Finally, in Fig. \ref{fig:comparison FID sm-abinitio A+-20} we compare the FID curves from Fig. 3a with similar ones obtained by introducing a 20\% modification of the hyperfine tensor due to a slight core spin polarization correction. We see that the qualitative effect on the absolute value of the FID coherence is marginal. Consequently, it is not observable in the Hahn-echo process which is shielded from the nuclear spin bath.

\section{\label{app:B}Induction considerations}

In this section we give a non-rigorous derivation of Eq. \ref{eq:fidcohsum} of the main text, starting from Eq. \ref{eq:fidcohprod}, by exploiting induction considerations. We consider the case where there is a single nuclear spin in our bath, i.e. $n=1$. In this case, by using trigonometric formulas, Eq. \ref{eq:fidcohprod} can be written as

\begin{widetext}
\begin{equation}
\Sigma_{1}(\tau)=\frac{1}{2}\left(S_{I_{1}}\cos\left[\left(\frac{\omega_{I_{1}}}{2}-\frac{\Omega_{I_{1}}}{2}\right)\tau\right]+D_{I_{1}}\cos\left[\left(\frac{\omega_{I_{1}}}{2}+\frac{\Omega_{I_{1}}}{2}\right)\tau\right]\right), \label{eq:fidcohprodn1}    
\end{equation}
\end{widetext}

\noindent\normalsize{and} thus we have two terms and two modulation frequencies, one for each term. In the case $n=2$ we have instead four terms and eight modulation frequencies, two for each term, as it can be seen by the following formula:

\begin{figure}
\centering
\begin{subfigure}{0.4944\linewidth}
\caption{}
\label{fig:var nbaths FID sm 500G}
    \includegraphics[width=0.8579\linewidth]{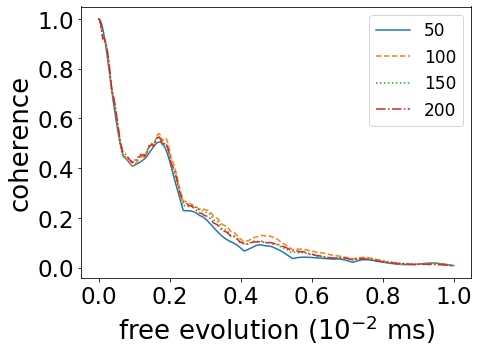}
\end{subfigure}
\hfill
\begin{subfigure}{0.4944\linewidth}
\caption{}
\label{fig:var nbaths FID abinit 500G}
    \includegraphics[width=0.8579\linewidth]{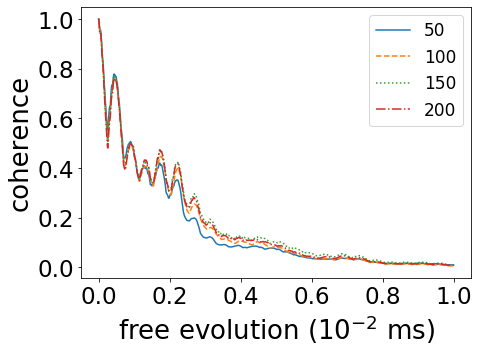}
\end{subfigure}
\caption{CCE1 absolute value of the coherence of a neutral $\mathrm{V}_{\mathrm{Si}}$ in 3C-SiC as a function of free evolution time, with different values for the number of bath realizations $\mathcal{N}$, for an external magnetic field of 500 G. The dynamics implemented is a free evolution, or FID process, and the radius of the spherical bath is $R_{\mathrm{bath}}=4\hspace{0.1cm}\mathrm{nm}$. (a): the hyperfine tensor components are calculated via Eq. \ref{eq:smhyptens} of the main text. (b): the hyperfine tensor components are calculated via \textit{ab initio} methods based on DFT (Table \ref{tab:my_label} of the main text).}
\label{fig:two graphs var nbaths FID}
\end{figure}

\footnotesize{
\begin{widetext}
\begin{equation} \label{eq:fidcohprodn2}
\begin{split}
\Sigma_{2}&=\frac{1}{8}\Bigg\lbrace S_{I_{1}}S_{I_{2}}\bigg[\cos\Big[\Big(\frac{\omega_{I_{1}}}{2}-\frac{\Omega_{I_{1}}}{2}+\frac{\omega_{I_{2}}}{2}-\frac{\Omega_{I_{2}}}{2}\Big)\tau\Big]+\cos\Big[\Big(\frac{\omega_{I_{1}}}{2}-\frac{\Omega_{I_{1}}}{2}-\frac{\omega_{I_{2}}}{2}+\frac{\Omega_{I_{2}}}{2}\Big)\tau\Big]\bigg]\\
&+S_{I_{1}}D_{I_{2}}\bigg[\cos\Big[\Big(\frac{\omega_{I_{1}}}{2}-\frac{\Omega_{I_{1}}}{2}+\frac{\omega_{I_{2}}}{2}+\frac{\Omega_{I_{2}}}{2}\Big)\tau\Big]+\cos\Big[\Big(\frac{\omega_{I_{1}}}{2}-\frac{\Omega_{I_{1}}}{2}-\frac{\omega_{I_{2}}}{2}-\frac{\Omega_{I_{2}}}{2}\Big)\tau\Big]\bigg]\\
&+D_{I_{1}}S_{I_{2}}\bigg[\cos\Big[\Big(\frac{\omega_{I_{1}}}{2}+\frac{\Omega_{I_{1}}}{2}+\frac{\omega_{I_{2}}}{2}-\frac{\Omega_{I_{2}}}{2}\Big)\tau\Big]+\cos\Big[\Big(\frac{\omega_{I_{1}}}{2}+\frac{\Omega_{I_{1}}}{2}-\frac{\omega_{I_{2}}}{2}+\frac{\Omega_{I_{2}}}{2}\Big)\tau\Big]\bigg]\\
&+D_{I_{1}}D_{I_{2}}\bigg[\cos\Big[\Big(\frac{\omega_{I_{1}}}{2}+\frac{\Omega_{I_{1}}}{2}+\frac{\omega_{I_{2}}}{2}+\frac{\Omega_{I_{2}}}{2}\Big)\tau\Big]+\cos\Big[\Big(\frac{\omega_{I_{1}}}{2}+\frac{\Omega_{I_{1}}}{2}-\frac{\omega_{I_{2}}}{2}-\frac{\Omega_{I_{2}}}{2}\Big)\tau\Big]\bigg]\Bigg\rbrace.
\end{split}
\end{equation}
\end{widetext}}

\noindent \normalsize{By} analyzing Eqs. \ref{eq:fidcohprodn1} and \ref{eq:fidcohprodn2} we find some common behavior that allows to infer the form of the equation valid in the general $n=N$ case to be exactly Eq. \ref{eq:fidcohsum}. Furthermore, in the general case we have $2^N$ terms and $2^{2N-1}$ modulation frequencies, $2^{N-1}$ for each term. The number of terms and the number of frequencies per term are not random, and can be understood, or counted, as the number of ways in which we can dispose $N$ elements from a set of $2$ elements, where the same element can be repeated at most $N$ times (they are thus called dispositions with repetitions). For what concerns the number of terms we have to dispose $N$ elements from the set of values $\lbrace S_{I_{i}},D_{I_{i}}\rbrace$ they can take, with a maximum of $N$ possible repetitions. The number of these dispositions is precisely $2^N$. Instead, for the number of frequencies per term we have to dispose $N-1$ \textit{pairs} of elements (pairs of signs), all but the first one, from the set of values $\lbrace+,-\rbrace$ they can take, with a maximum of $N-1$ possible repetitions, \textit{fixing} at the same time the first pair to a $(+-)$ if the term they are multiplied by starts with an $S$ or to a $(++)$ if the term they are multiplied by starts with a $D$. Since the cosine is an even function, we can make the opposite choice, $S\longrightarrow(-+)$ and $D\longrightarrow(--)$, but also in this case the rules remain the same and nothing changes. \\
\hspace*{0.5cm}Finally, we give the version of Eq. \ref{eq:fidcohimpartsum} in which also the $\cos(\omega_{1}\tau)$ term is put inside of $\Sigma_{N}$, i.e.

\begin{widetext}
\begin{equation} \label{eq:fidcohimpartfinal}
\begin{split}
\langle S_{y}\rangle&=\frac{1}{2^{2N}}\bigg\lbrace S_{I_{1}}\cdots S_{I_{N}}\Big[(-+-+\ldots-+)+\ldots+(-++-\ldots+-)\\
&+(+--+\ldots-+)+\ldots+(+-+-\ldots+-)\Big]\\
&+S_{I_{1}}\cdots D_{I_{N}}\Big[(-+\ldots-+--)+\ldots+(+-\ldots+---)\\
&+(-+\ldots-+++)+\ldots+(+-\ldots+-++)\Big]+\ldots\\
&+S_{I_{1}}\cdots D_{I_{i}}\cdots D_{I_{N}}\Big[(-+\ldots--\ldots--)+\ldots+(+-\ldots--\ldots++)\\
&+(-+\ldots++\ldots--)+\ldots+(+-\ldots++\ldots++)\Big]+\ldots\\
&+D_{I_{1}}\cdots D_{I_{N}}\Big[(----\ldots--)+\ldots+(--++\ldots++)\\
&+(++--\ldots--)+\ldots+(++++\ldots++)\Big]\bigg\rbrace,    
\end{split}    
\end{equation}
\end{widetext}

\begin{figure}
\begin{center}
    \includegraphics[width=0.4\textwidth]{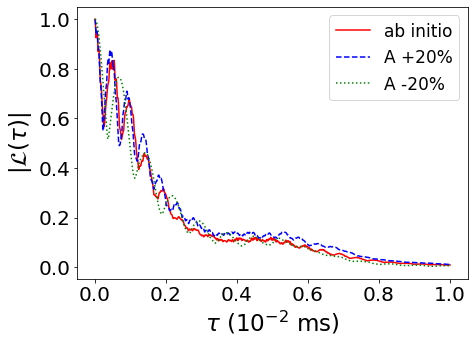}
    \caption{FID evaluated with CCE2 with semiclassical and \textit{ab initio} hyperfine tensor components: absolute value of the coherence of a neutral $\mathrm{V}_{\mathrm{Si}}$ in 3C-SiC as a function of free evolution time, for an external magnetic field of 200 G. The result is averaged over 50 different baths. In the dashed and dotted curves the hyperfine tensor has been varied by 20\%.}
    \label{fig:comparison FID sm-abinitio A+-20}
\end{center}
\end{figure}

\noindent where now a new notation is used,

\begin{equation}
\begin{split}
(+-\ldots+-)&\equiv\cos\big[(\omega_{1}+\omega_{I_{1}}/2-\Omega_{I_{1}}/2+\ldots \\
&+\omega_{I_{N}}/2-\Omega_{I_{N}}/2)\tau\big]. \label{eq:fidnewshortnot}
\end{split}
\end{equation}

\noindent In this case there are $2^{2N}$ modulation frequencies, thus their number is doubled in size, as noticed in the main text. Furthermore, each of the final modulation frequencies appearing in Eq. \ref{eq:fidcohimpartfinal} can be written as a linear combination of the eigenvalues of the pure-dephasing Hamiltonian, given in Eq. \ref{eq:hamtotpuredeph}, in the CCE1 case (see Appendix \ref{app:C}). \\
\hspace*{0.5cm}Now, an interesting calculation to perform is the one involving the modulation frequencies, which can be analytically obtained through Eq. \ref{eq:fidcohsum}, in a specific case. The Fourier transform of the signal in time, given in Eq. \ref{eq:fidcohsum}, is easily obtained and can be written as

\small{
\begin{widetext}
\begin{equation} \label{eq:fidcohsumtrans}
\begin{split}
\Tilde{\Sigma}_{N}(\omega)&=\frac{\pi}{2^{2N-1}}\bigg\lbrace S_{I_{1}}\cdots S_{I_{N}}\Big[[+-+-\ldots+-]+[-+-+\ldots-+]+\ldots\\
&+[+--+\ldots-+]+[-++-\ldots+-]\Big]\\
&+S_{I_{1}}\cdots D_{I_{N}}\Big[[+-+-\ldots++]+[-+-+\ldots--]+\ldots\\
&+[+--+\ldots--]+[-++-\ldots++]\Big]+\ldots\\
&+S_{I_{1}}\cdots D_{I_{i}}\cdots D_{I_{N}}\Big[[+-\ldots++\ldots++]+[-+\ldots--\ldots--]+\ldots\\
&+[+-\ldots--\ldots--]+[-+\ldots++\ldots++]\Big]\\
&+\ldots+D_{I_{1}}\cdots D_{I_{N}}\Big[[++++\ldots++]+[----\ldots--]+\ldots\\
&+[++--\ldots--]+[--++\ldots++]\Big]\bigg\rbrace,
\end{split}
\end{equation}
\end{widetext}}\noindent\normalsize{where}
now, each term inside the curly brackets is multiplied by a sum of Dirac delta functions, and we have introduced the notation

\begin{equation}
\begin{split}
[+-\ldots+-]&\equiv\delta\big[\omega+\omega_{I_{1}}/2-\Omega_{I_{1}}/2+\ldots \\
&+\omega_{I_{N}}/2-\Omega_{I_{N}}/2\big]. \label{eq:fidshortnottrans}
\end{split}
\end{equation}

\noindent In passing from Eq. \ref{eq:fidcohsum} to Eq. \ref{eq:fidcohsumtrans}, we have used the known result

\begin{equation}
\mathcal{F}[\cos(\omega_{0}t)]=\pi(\delta[\omega+\omega_{0}]+\delta[\omega-\omega_{0}]). \label{eq:cosfouriertrans}    
\end{equation}

\begin{figure}
\centering
\begin{subfigure}{0.4944\linewidth}
\caption{}
\label{fig:FIDsig_n1}
    \includegraphics[width=1.0\linewidth]{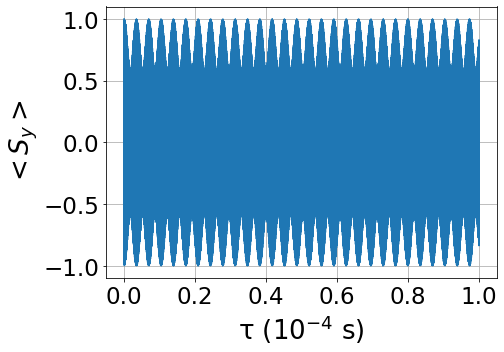}
\end{subfigure}
\hfill
\begin{subfigure}{0.4944\linewidth}
\caption{}
\label{fig:Fourier_trans_n1}
    \includegraphics[width=1.0\linewidth]{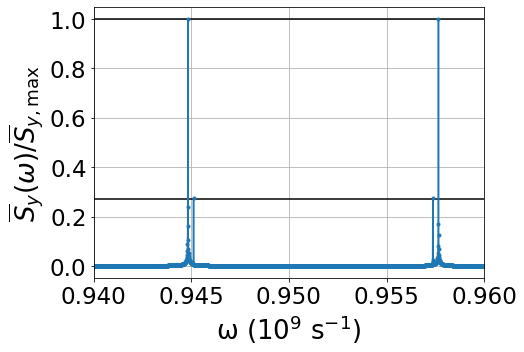}
\end{subfigure}
\caption{FID evaluated with CCE1 with \textit{ab initio} hyperfine tensor components: (a) imaginary part of the coherence of a neutral $\mathrm{V}_{\mathrm{Si}}$ in 3C-SiC as a function of free evolution time, for an external magnetic field of 340 G. The bath contains a single $^{29}\mathrm{Si}$ nucleus in the second-neighbor shell. (b) Normalized Fourier transform of the signal in the time domain obtained via a numerical FFT algorithm.}
\label{fig:num_Fourier_trans_n1}
\end{figure}

\noindent By examining the signal in frequency, given in Eq. \ref{eq:fidcohsumtrans}, we see that the Fourier transform will display delta-like peaks in correspondence to each modulation frequency inside the arguments of the delta functions. To exemplify this by means of an example, we report in Fig. \ref{fig:num_Fourier_trans_n1} the imaginary part of the FID time signal of a $\mathrm{V}_{\mathrm{Si}}^0$ interacting with a single $^{29}\mathrm{Si}$ nucleus in the second-neighbor shell, for simplicity, and its normalized Fourier transform obtained via a numerical Fast Fourier Transform (FFT) algorithm. Since in this case $N=1$, the modulation frequencies are $4$ and are quantitatively predicted by Eq. \ref{eq:fidcohimpartfinal} to be $0.9448\hspace{0.1cm}\mathrm{GHz}$, $0.9576\hspace{0.1cm}\mathrm{GHz}$, $0.9574\hspace{0.1cm}\mathrm{GHz}$ and $0.9451\hspace{0.1cm}\mathrm{GHz}$, respectively. The relative amplitude of the central peaks is given by $(1-\frac{\omega_{I}+A}{\Omega_{I}})/(1+\frac{\omega_{I}+A}{\Omega_{I}})=0.27$ (see Fig. \ref{fig:Fourier_trans_n1}). Even though Eq. \ref{eq:fidcohimpartfinal} can be used to find the modulation frequencies displayed in Fig. 3, a graphical depiction of the peaks of the corresponding Fourier transform as in Fig. \ref{fig:Fourier_trans_n1} is impractical due to the presence of $\sim2^{3000}$ of them for an entire nuclear spin bath. Furthermore, most of them would not be visible due to the reduced resolution necessary to in principle show all of them in the horizontal axis and the ever-decreasing relative amplitudes of the central peaks. In this sense, the example in Fig. \ref{fig:num_Fourier_trans_n1} is extremely useful as a proof of principle of Eq. \ref{eq:fidcohimpartfinal}, in that it shows all the frequencies predicted as peaks in the corresponding Fourier transform in a simple case. Once we have demonstrated the reliability of Eq. \ref{eq:fidcohimpartfinal}, through the example in Fig. \ref{fig:num_Fourier_trans_n1} and how we have obtained it, we can proceed towards its application to finding the dominant frequencies of the curves in Fig. 3. We already know that the dominant frequencies, the ones represented by peaks with unitary amplitude in the FFT, are multiplied by the term $SS\cdots S$ (the first term), since after normalization they are multiplied by 1 and the others by a number between 0 and 1. As we detail in this section, these frequencies are $2^N$ ($\sim2^{1500}$ for a full bath). One of them, the others being very close, differing in the second decimal place (see Fig. \ref{fig:Fourier_trans_n1}), is contained in the term $(+-+-\ldots+-)$, i.e.

\begin{equation}
\begin{split}
\omega_{\mathrm{dom}}&=\omega_{1}+\omega_{I_{1}}/2-\Omega_{I_{1}}/2+\omega_{I_{2}}/2-\Omega_{I_{2}}/2+\ldots \\
&+\omega_{I_{N}}/2-\Omega_{I_{N}}/2. \label{eq:omegadom}
\end{split}
\end{equation}

\section{\label{app:C}Modulation frequencies}

\begin{figure*}
\begin{center}
    \includegraphics[width=1.0\textwidth]{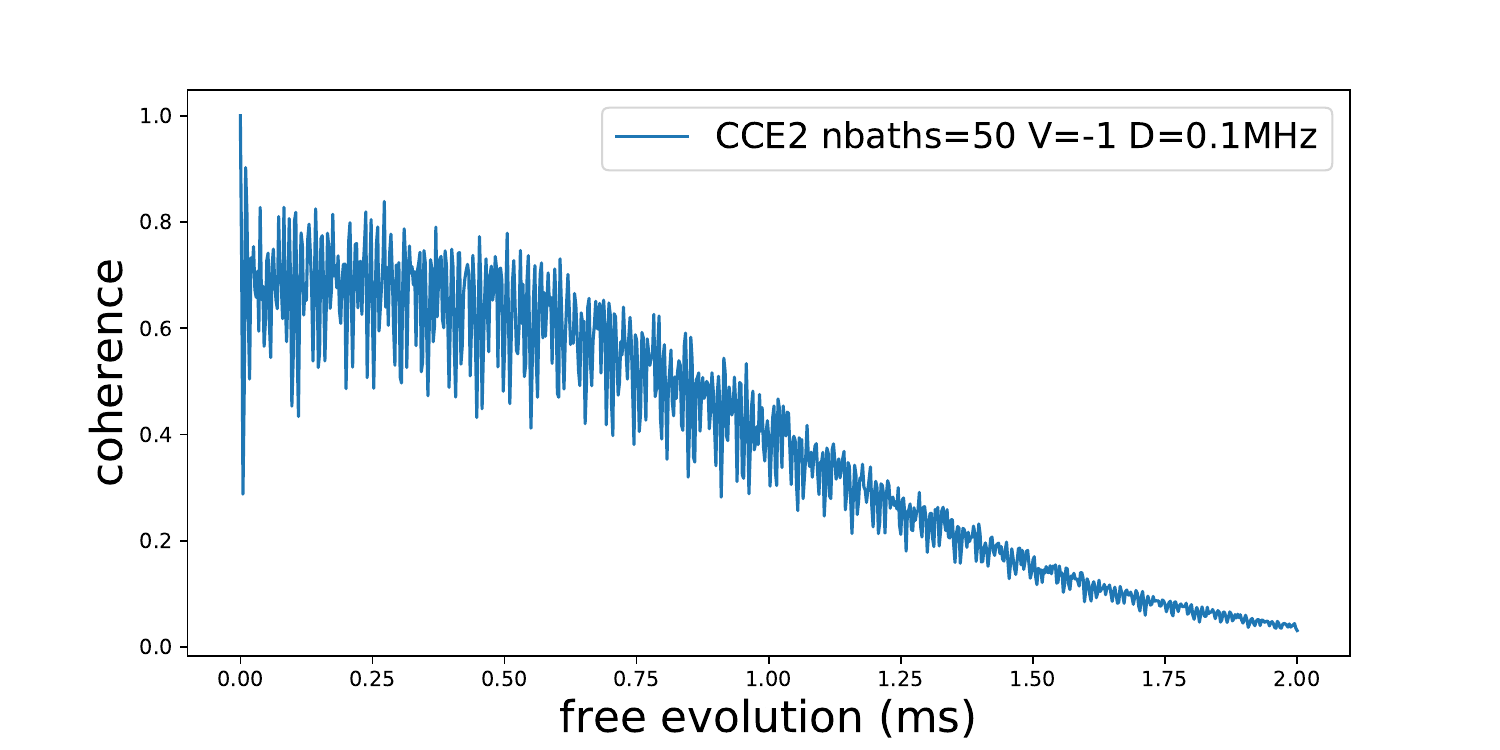}
    \caption{Hahn-echo evaluated with CCE2 \textit{ab initio} hyperfine tensor components (see Table \ref{tab:my_label_1} in the main text): absolute value of the coherence of a negatively charged $\mathrm{V}_{\mathrm{Si}}$ in 3C-SiC as a function of free evolution time, for an external magnetic field of 200 G. The result is averaged over 50 different baths.}
    \label{fig:CEE_Vneg}
\end{center}
\end{figure*}

In this section we derive the modulation frequencies from the FID subsection of the main text by following a different approach. We start with the case of only one nuclear spin in the bath, i.e. $n=1$, and write the pure-dephasing Hamiltonian \ref{eq:hamtotpuredeph} in this case,

\vspace{0.5em}
\begin{equation}
\mathcal{H}=D S_{z}^2+\gamma_{e}B_{z}S_{z}+\gamma_{1}B_{z}S_{z}+A S_{z}I_{z}+B S_{z}I_{x}. \label{eq:hamtotpuredephn1}    
\end{equation}
\noindent The problem's Hilbert space has dimension $\mathrm{dim}\left(\mathscr{H}\right)=3\times2=6$, so that the matrix representing Hamiltonian \ref{eq:hamtotpuredephn1} is a $6\times6$ one. We report the matrix in the following:
\begin{widetext}
\begin{equation}
\begin{pmatrix}
\omega_{1}+\frac{\omega_{I}}{2}+\frac{A}{2} & B/2 & 0 & 0 & 0 & 0\\
B/2 & \omega_{1}-\frac{\omega_{I}}{2}-\frac{A}{2} & 0 & 0 & 0 & 0\\
0 & 0 & \frac{\omega_{I}}{2} & 0 & 0 & 0\\
0 & 0 & 0 & -\frac{\omega_{I}}{2} & 0 & 0\\
0 & 0 & 0 & 0 & \omega_{-1}+\frac{\omega_{I}}{2}-\frac{A}{2} & -B/2\\
0 & 0 & 0 & 0 & -B/2 & \omega_{-1}-\frac{\omega_{I}}{2}+\frac{A}{2} 
\end{pmatrix}.
\end{equation}
\end{widetext}

\noindent This is a block-diagonal matrix with three quadrants, one for each of the three electron spin's energy levels. Consequently, the third quadrant is of no interest for us, whereas the second one is already diagonal. It is required to diagonalize the first quadrant to be able to write down the first four eigenvalues, which are given by
\begin{align}
\omega_{1}+\frac{\Omega_{I}}{2}&=E_{1}, \label{eq:eig1}\\
\omega_{1}-\frac{\Omega_{I}}{2}&=E_{2}, \label{eq:eig2}\\
\frac{\omega_{I}}{2}&=E_{3}, \label{eq:eig3}\\
-\frac{\omega_{I}}{2}&=E_{4}, \label{eq:eig4}
\end{align}
where $\Omega_{I}$ is defined in Eq. \ref{eq:OmegaIi} of the main text. Now, from Eq. \ref{eq:fidcohimpartfinal} we have the $2^{2N}$, in this case $4$, modulation frequencies, which are the following,
\begin{align}
\omega_{1}-\frac{\omega_{I}}{2}+\frac{\Omega_{I}}{2}&=E_{1}+E_{4}, \label{eq:relmodfreqeig1}\\
\omega_{1}+\frac{\omega_{I}}{2}-\frac{\Omega_{I}}{2}&=E_{2}+E_{3}, \label{eq:relmodfreqeig2}\\
\omega_{1}-\frac{\omega_{I}}{2}-\frac{\Omega_{I}}{2}&=E_{2}+E_{4}, \label{eq:relmodfreqeig3}\\
\omega_{1}+\frac{\omega_{I}}{2}+\frac{\Omega_{I}}{2}&=E_{1}+E_{3}. \label{eq:relmodfreqeig4}
\end{align}
Eqs. \ref{eq:relmodfreqeig1}-\ref{eq:relmodfreqeig4} are the relations regarding the modulation frequencies and our system's eigenenergies. Furthermore, analogous relations valid in the case $n=N$ can be easily obtained by using induction considerations.

\section{\label{app:D}CCE2 based estimate of $T_2$ for negatively charged Si vacancy in 3C-SiC}

Here we evaluate the coherence decay for a negatively charged $\mathrm{V}_{\mathrm{Si}}$ in 3C-SiC by applying the same Hahn-echo protocol discussed in subsection \ref{subsec:hahn-echo} of the main text. In this case the initial state dynamics is a superposition of the $\lbrace\vert1/2\rangle,\vert3/2\rangle\rbrace$ eigenstates of the $S_{z}$ spin operator 
\begin{equation}
\vert\Psi\rangle=\frac{1}{\sqrt{2}}\big(\vert3/2 \rangle+i\vert1/2\rangle\big), \label{eq:initstate3/2}
\end{equation}
and the qubit dynamics occur in the corresponding  subset of the $S_{z}$.
Fig. \ref{fig:CEE_Vneg} shows the qubit coherence calculated at the CCE2 level of the theory as a function of free evolution time for an external magnetic field of 200 G. We used the ab initio estimates for the hyperfine tensor components for nuclear spins in the first and second neighbor shells reported in Table II of the main text, which are different with respect to the same quantities calculated for the neutral defect.
In addition, we notice in this case that due to the two $S_{z}$ eigenvalues which are not null, the spin modulation is ruled by two shifted frequencies with respect to the Larmor one \cite{nagy2019high}. Therefore, the dynamics and the interactions with the nuclear spins are intrinsically different with respect to the $S=1$ neutral defect. As a consequence, the coherence decay seems to show two time components. Anyway, if we extract the slower one with a stretched exponential function we can estimate optimal values of $T_{2}=1.16\hspace{0.1cm}\mathrm{ms}$ and $n=2.05$, respectively which are very similar to the ones derived for the $\mathrm{V}_{\mathrm{Si}}^{0}$ defect.

%\section*{Author contributions statement}

%T.F. did the conceptualization (lead), data curation (equal), formal analysis (equal), investigation (equal), methodology (equal), software (equal), visualization (equal) and writing (equal). G.F. dealt with the investigation (equal) and validation (equal). I.D. was involved in the investigation (equal), methodology (equal), supervision (equal) and validation (equal). E.P. was involved in the formal analysis (equal), investigation (equal), methodology (equal), supervision (equal) and validation (equal). A.L.M. was involved in the formal analysis (equal), investigation (equal), methodology (equal), supervision (equal) and validation (equal). All authors reviewed the manuscript.

%\section*{Data availability statement}

%Some of the data are available under reasonable request to the corresponding author. 

%\section*{Additional information}

%T.F. acknowledges no conflict of interests on behalf of all authors of the paper.

%\nocite{*}

\bibliography{apssamp}% Produces the bibliography via BibTeX.

\end{document}